\documentclass[a4paper,fleqn,usenatbib]{mnras}

\usepackage{mathptmx}

\usepackage[T1]{fontenc}
\usepackage{ae,aecompl}

\usepackage{graphicx}	
\usepackage{amsmath}	
\usepackage{amssymb}	
\usepackage{setspace}

\newcommand{\xmm}{\textit{XMM-Newton}}

\newcommand{\swift}{\textit{Swift}}
\newcommand{\delcstat}{$\Delta$C-stat}

\title[Her X-1 accretion disc wind]{An ionised accretion disc wind in Hercules X-1}

\author[P Kosec et al.]{P. Kosec$^{1}$\thanks{E-mail: pk394@cam.ac.uk}, 
A. C. Fabian$^{1}$, 
C. Pinto$^{2}$,
D. J. Walton$^{1}$,
S. Dyda$^{1}$ \newauthor
and C. S. Reynolds$^{1}$
\\
$^{1}$Institute of Astronomy, Madingley Road, CB3 0HA Cambridge, UK \\
$^{2}$ESTEC/ESA, Keplerlaan 1, 2201AZ Noordwijk, The Netherlands \\
}

\date{Accepted 2019 November 11. Received 2019 October 28; in original form 2019 August 9}

\pubyear{2019}

\begin{document}
\label{firstpage}
\pagerange{\pageref{firstpage}--\pageref{lastpage}}
\maketitle

\begin{abstract}

Hercules X-1 is one of the best studied highly magnetised neutron star X-ray binaries with a wealth of archival data. We present the discovery of an ionised wind in its X-ray spectrum when the source is in the high state. The wind detection is statistically significant in most of the \xmm\ observations, with velocities ranging from 200 to 1000 km/s. Observed features in the iron K band can be explained by both wind absorption or by a forest of iron emission lines. However, we also detect nitrogen, oxygen and neon absorption lines at the same systematic velocity in the high-resolution RGS grating spectra. The wind must be launched from the accretion disc, and could be the progenitor of the UV absorption features observed at comparable velocities, but the latter likely originate at significantly larger distances from the compact object. We find strong correlations between the ionisation level of the outflowing material and the ionising luminosity as well as the super-orbital phase. If the luminosity is driving the correlation, the wind could be launched by a combination of Compton heating and radiation pressure. If instead the super-orbital phase is the driver for the variations, the observations are likely scanning the wind at different heights above the warped accretion disc. If this is the case, we can estimate the wind mass outflow rate, corrected for the limited launching solid angle, to be roughly 70\% of the mass accretion rate.

\end{abstract}




\section{Introduction}

Highly ionised winds have been well established in the X-ray spectra of a number of stellar mass accretors such as black hole \citep[e.g.][]{Miller+06, Neilsen+09} and neutron star binaries \citep[e.g.][]{Ueda+04, Miller+11}, with velocities of 100s to 1000s km/s. Such winds might in fact be a universal phenomenon during the bright soft state of black hole binaries \citep{Ponti+12}. The high ionisation state of these winds necessarily means that they are best observed in the Fe K energy band ($\sim7$ keV) located within the X-ray part of the electromagnetic spectrum. 

The origin of these outflows is the accretion disc of the compact object, but their driving mechanism and energy budget are currently still in question. As the wind material is highly ionised (usually $\xi>10^3$), line driving is unlikely to provide enough driving force \citep{Proga+00}, and since the objects in question are accreting below the Eddington limit, the radiation pressure on electrons is likely not sufficient to launch significant winds \citep[although see][]{Neilsen+16}. If the base of the wind is illuminated by the hard X-ray radiation from the inner accretion flow, the material could be Compton heated and launched in form of a thermally-driven wind \citep{Begelman+83}. This effect was invoked to explain the winds in a number of sources \citep{Neilsen+09, DiazTrigo+12}. Alternatively, the wind could be driven by magnetic forces \citep{Miller+06, Miller+08, Fukumura+17}.

Evidence of even faster winds ($\sim$0.2c) has been found in Ultraluminous X-ray sources \citep{Pinto+16, Pinto+17, Walton+16}, which are thought to be powered by super-Eddington stellar mass accretors, including one ULX harbouring a neutron star \citep{Kosec+18b}. In these objects, radiation pressure might be the natural driving mechanism of the outflowing material. How these winds in stellar-mass accretors relate to the ultrafast outflows \citep[e.g.][]{Pounds+03, Reeves+03, Parker+16} and warm absorbers \citep[e.g.][]{Reynolds+95, Lee+01} observed in active galactic nuclei and tidal disruption events \citep{Miller+15a, Kara+18} and whether they are driven by the same phenomenon, is currently not understood. It is, however, certain that accretion disc winds play a major role in the phenomenon of accretion and thus need to be studied in detail.

Here we present the discovery of highly ionised blueshifted absorption in the X-ray spectrum of the neutron star X-ray binary Hercules X-1 in the high state. Previous work found blueshifted UV absorption lines attributed to a circumbinary wind launched from the irradiated surface of the secondary \citep{Boroson+01}, but only weak signatures ($\sim2\sigma$) of an X-ray counterpart were found so far \citep{Leahy+19}. We significantly detect the outflowing material in most of the high state observations made with \xmm\ (using RGS and pn instruments), at a velocity of 200-1000 km/s. We find that the wind is launched from within the accretion disc of the primary, and that the mass outflow rate is of the same order as the mass accretion rate onto the neutron star. We conclude that the wind is most likely driven by magnetic fields or by Compton heating, but at the moment we cannot pinpoint the exact mechanism.

\subsection{Hercules X-1}

Hercules X-1 \citep[hereafter Her X-1,][]{Giacconi+72} is one of the most famous, brightest and most studied neutron star X-ray binaries. The system consists of a highly magnetised neutron star and a 2 $M_{\odot}$ secondary HZ Herculis \citep{Middleditch+76}. Her X-1 is especially well known for the three different timescales of multiwavelength variability. 

The shortest is the 1.24 s rotation period \citep{Giacconi+73} of the neutron star. The neutron star is known to harbour a strong magnetic field of the order of $10^{12}$ G, manifested by a cyclotron scattering feature with an energy of 35-40 keV \citep{Truemper+78, Staubert+07}. 

The second important timescale is the 1.7-day orbital period of the binary, accompanied by X-ray eclipses \citep{Bahcall+72}, suggesting that the system is observed almost edge-on. The longest timescale is the 35-day super-orbital cycle of high, low and short-on X-ray flux states \citep{Tananbaum+72}. Each cycle begins by a 10 day high state with a brightness of $\sim 4 \times 10^{37}$ erg/s, followed by a low state during which the flux drops by a factor of 10. This is followed by a short-on state \citep{Fabian+73} with a flux of $\sim$1/3 of the maximum flux for a few days and then again by a low state.

Such behaviour can be explained by a model according to which the accretion disc of the neutron star (seen almost edge-on) is warped \citep{Ogilvie+01} and precesses with a 35-day period \citep{Gerend+76}. Although accretion onto the compact object continues at a steady rate, inner parts of the disc and the object itself are obscured from our view for extended parts of the cycle. The hard X-ray radiation originates in the accretion column \citep{Ghosh+79} near the surface of the neutron star and is beamed \citep{Scott+00}. The size of the magnetosphere \citep{Lamb+73} of $\sim2 \times 10^8$ cm $\approx 1000~$R$_{\textrm{G}}$ defines the inner edge of the accretion disc. The outer edge of the disc is at about $2 \times 10^{11}$ cm ($10^{6}~$R$_{\textrm{G}}$) and the binary separation is around $3 \times 10^{11}$ cm \citep{Cheng+95}.

Throughout the manuscript, we adopt a distance of Her X-1 of 6.1 kpc \citep{Leahy+14}. All the uncertainties are stated at $1\sigma$ level.

\section{Observations and Data Reduction}

\begin{table*}
\centering
\caption{Log of the observations used in this work. Observational IDs as well as start dates are listed in the first 2 columns. The following columns contain the clean exposures of RGS1, RGS2 and pn instruments for each observation as well as the average count rates. The last column contains following notes about individual observations: 1 - RGS 2 instrument was not operational during this observation, 2 - RGS 1 instrument was not operational during this observation, 3 - obvious dips were visible in the lightcurve during the observation, only high flux periods were extracted for this analysis.}
\label{Obsdata}
\begin{tabular}{ccccccccc} 
Obs. ID&Obs. date&\multicolumn{3}{|c|}{Exposures (s)}&\multicolumn{3}{|c|}{Average count rates (s$^{-1}$)}&Notes\\
&&RGS1&RGS2&pn&RGS1&RGS2&pn&\\

\hline
0134120101&2001-01-26&11296&------&5659.1&11.5&------&389.3&1\\
0153950301&2002-03-17&------&7257.9&2773.6&------&19.2&591.8&2\\
0673510501&2011-07-31&8375.5&8361.9&6802.8&11.5&12.6&456.0&3\\
0673510601&2011-09-07&32242&32130&19320&16.5&17.8&706.6&\\
0673510801&2012-02-28&12801&12741&4815.2&19.4&20.9&794.1&\\
0673510901&2012-04-01&13122&13029&9425.6&11.6&12.7&491.8&\\
0783770501&2016-08-17&4597.8&4550.3&4588.4&4.03&4.37&206.8&3\\
0783770601&2016-08-17&5233.9&5222.9&4492.0&9.38&10.3&399.4&3\\
0783770701&2016-08-18&12331 &12277 &6885.9&12.9&14.2&554.7&\\

\hline
\end{tabular}
\end{table*}

\xmm\ \citep{Jansen+01} data were used in this study as the observatory offers a combination of good collecting area as well as very good spectral resolution. Furthermore, its archive contains a wealth of data on Her X-1. Initially, we utilise all \xmm\ observations with good enough statistics for analysis of absorption lines in the RGS grating data. Considering the current archive of Her X-1, this effectively limits us to observations of the object in the high state. Low and short-on state grating observations are individually sufficient for emission line studies \citep{Jimenez-Garate+02}, however the continuum flux is too low for absorption line searches. All of the high state observations of Her X-1 used in this study are listed in Table \ref{Obsdata} along with their clean exposures and count rate information. Whenever possible, we use simultaneously Reflection Grating Spectrometer \citep[RGS,][]{denHerder+01} data which offer best spectral resolution, as well as European Photon Imaging Camera (EPIC) pn \citep{Struder+01} data to capture the broadband continuum outside the RGS band. All the available archival \xmm\ data were downloaded from the XSA archive and reduced using a standard pipeline with SAS v16, CalDB as of July 2017.

RGS data were extracted using standard routines with default source and background selection regions. High-background periods were filtered with a threshold of 0.25 counts/sec. Both first- and second-order data were extracted but as the count rate in the second order is much lower, it was only used in selected instances. The second order RGS 1 spectrum was used during the analysis of observation 0134120101 where RGS 2 detector was not working, and conversely the second order RGS 2 spectrum was used in observation 0153950301 where RGS 1 was not operational. The first-order data were binned by a factor of 3 directly in the SPEX fitting package to oversample the real grating resolution by about a factor of 3. The second order data were binned by a factor of 6 to achieve similar spectral binning as the first order data. The data were used in the spectral range between 7 \AA\ (1.8 keV) and 35 \AA\ (0.35 keV).

The EPIC pn detector was in the Timing mode during all observations due to very high Her X-1 flux. The calibration accuracy of the Timing mode can be found in the following \xmm\ Calibration Technical Note\footnote{http://xmm2.esac.esa.int/docs/documents/CAL-TN-0083.pdf}. High-background periods were filtered on case-by-case basis as the standard routine thresholds were often below the actual Her X-1 fluxes in the high state. Events of PATTERN <= 4 (single/double) were accepted for pn data, and the background regions were very small rectangles as far from the source position as possible. However, the background was very weak (usually $<3$\% of source flux)  compared to the source flux in all of the observations used in this study. The data were grouped to at least 25 counts per bin and also binned by at least a factor 3 using the SPECGROUP procedure. pn data were used in the spectral range of 3 keV to 10 keV. Data below 2 keV were not necessary as this range was covered by the RGS instrument with superior spectral resolution. The range between 2 and 3 keV contained in multiple instances strong residuals and was ignored in this study. These residuals were also present in lower flux observations and therefore were not likely caused only by pile-up. They could be a result of incorrect calibration around the Au edge at 2.3 keV \citep{Pintore+14}.

All the reduced data were converted from the OGIP format to SPEX format using the Trafo tool. We use the SPEX fitting package \citep{Kaastra+96} for spectral fitting as it contains several useful ionised absorption models such as \textsc{pion} and \textsc{xabs}. Both models calculate the absorption spectrum from physical absorber parameters on the fly and therefore do not require any grid preparation before data fitting (as is necessary with \textsc{xstar}). All the data were fitted with Cash statistic \citep{Cash+76}.

\subsection{Orbital and super-orbital phases}

We determine the exact orbital and super-orbital phase of the system for each of the \xmm\ observations. The phases are listed in Table \ref{ObsPhases}.

The orbital phase of each observation was determined using \xmm\ GTI files. The starting and ending time of the exposure (in \xmm\ seconds) was extracted and converted to the Modified Julian Date (MJD). We then used the Her X-1 orbital solution from \citet{Staubert+09} to determine the current orbital phase. The error on the MJD and the orbital phase value listed in Table \ref{ObsPhases} is defined as half the clean exposure time of the observation. All the \xmm\ exposures are much shorter than the binary orbital period and therefore the orbital phase change during a single observation is relatively small.

The super-orbital phase of each observation was calculated by determining the high state turn-on point for the current and the next super-orbital cycle. We used the Swift/BAT Hard X-ray Transient Monitor \citep{Krimm+13} to produce a lightcurve of Her X-1 between 2007 and present time. The observation times were determined and then the neighbouring turn-on times were estimated from the lightcurve. The phase was calculated as the time elapsed from the previous turn-on divided by the time difference between the neighbouring turn-on times \cite[the super-orbital period is not stable and varies between 33 and 37 days,][]{Leahy+10}. We assume an uncertainty of 0.5 day for the turn-on point determination, which gives our errorbar on the super-orbital phase for each observation. Finally, the super-orbital phases of observations 0134120101 and 0153950301 (which happened before \swift\ was launched) were taken from \citet{Leahy+10}.

\begin{table}
\centering
\caption{Modified Julian dates, orbital and super-orbital periods of Her X-1 for each of the high state \xmm\ observations. The first column lists the observational ID, the second contains the MJD and the third the orbital phase of the observation midpoint. The errorbar on the MJD and orbital phase denotes half the length of each observation. The last column contains the super-orbital phases for each observation, and its errorbar is given by the accuracy of the high state turn-on time determination.}
\label{ObsPhases}
\begin{tabular}{cccc} 
\hline
Obs. ID&Midpoint&Orbital&Super-orbital\\
&MJD&phase&phase\\
0134120101&$ 51935.078 \pm 0.042$ &$0.210 \pm 0.025$ &$0.170 \pm 0.015$\\
0153950301&$ 52350.068 \pm 0.050$ &$0.297 \pm 0.029$ &$0.036 \pm 0.014$\\
0673510501&$ 55773.362 \pm 0.055$ &$0.802 \pm 0.032$ &$0.028 \pm 0.015$\\
0673510601&$ 55811.511 \pm 0.185$ &$0.240 \pm 0.108$ &$0.123 \pm 0.015$\\
0673510801&$ 55985.142 \pm 0.073$ &$0.366 \pm 0.043$ &$0.104 \pm 0.015$\\
0673510901&$ 56018.912 \pm 0.075$ &$0.229 \pm 0.044$ &$0.090 \pm 0.015$\\
0783770501&$ 57617.323 \pm 0.043$ &$0.379 \pm 0.025$ &$0.003 \pm 0.014$\\
0783770601&$ 57617.853 \pm 0.031$ &$0.690 \pm 0.018$ &$0.018 \pm 0.014$\\
0783770701&$ 57618.803 \pm 0.070$ &$0.249 \pm 0.041$ &$0.045 \pm 0.014$\\
\hline
\end{tabular}
\end{table}

We use the orbital phase information of each observation to correct the measured wind velocity for the orbital motion of the neutron star and its accretion disc. The projected orbital velocity of the neutron star is $\textrm{v~sin~i} = 169.049 \pm 0.004 $~km s$^{-1}$ \citep{Deeter+81} and the systemic velocity of the system was measured to be ($-65 \pm 2$)~km~s$^{-1}$ \citep{Reynolds+97}. The eccentricity of the binary is very small: ($4.2 \pm 0.8)\times10^{-4}$ \citep{Staubert+09}. We can therefore correct the outflow velocity by assuming a sinusoid shape of the orbital velocity evolution with time, such as:

\begin{equation}
v_{\textrm{corr}}=\textrm{v}_{0}+65~\textrm{km~s}^{-1}+169~\textrm{km~s}^{-1} \times \sin (2\pi\phi_{\textrm{orb}})
\end{equation}

where v$_{0}$ is negative (spectral lines are blueshifted) and $\phi_{\textrm{orb}}$ is the midpoint orbital phase for each observation ($\phi_{\textrm{orb}}=0$ defines the center of the neutron star eclipse). All the wind velocity values in this manuscript including those in Table \ref{Results} are already corrected for the binary motion.

\section{Results}
\label{results}

\subsection{Continuum modelling}
\label{continuummodelling}

\begin{figure}
	\includegraphics[width=\columnwidth]{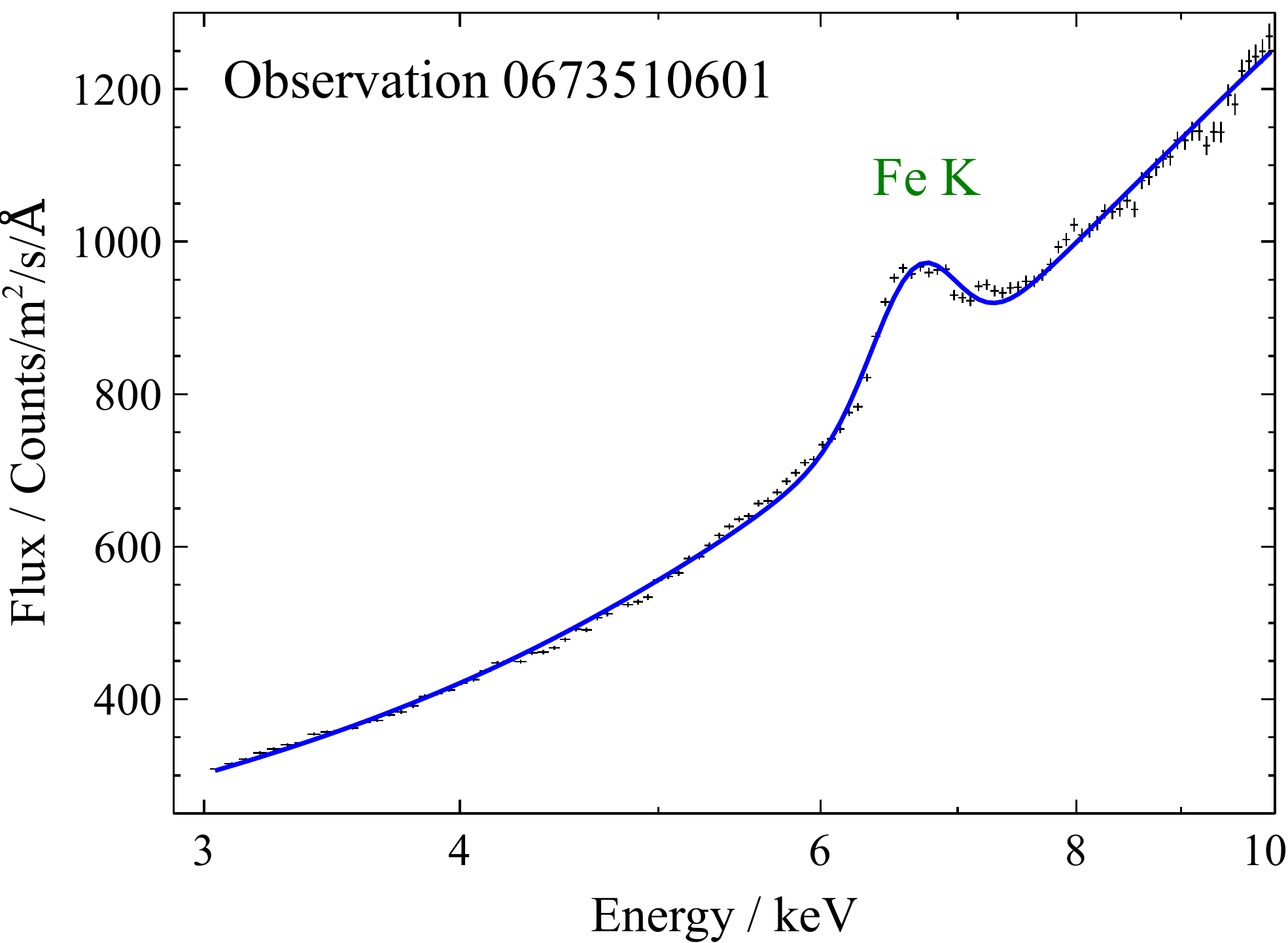}
    \caption{Example 3-10 keV spectrum of Her X-1 with \xmm\ pn instrument (Obs ID 0673510601). The broadband shape is reasonably well fitted with a Comptonisation model plus a broad iron line at $\sim$6.6 keV.}
    \label{Continuum_PN}
\end{figure}

\begin{figure*}
	\includegraphics[width=\textwidth]{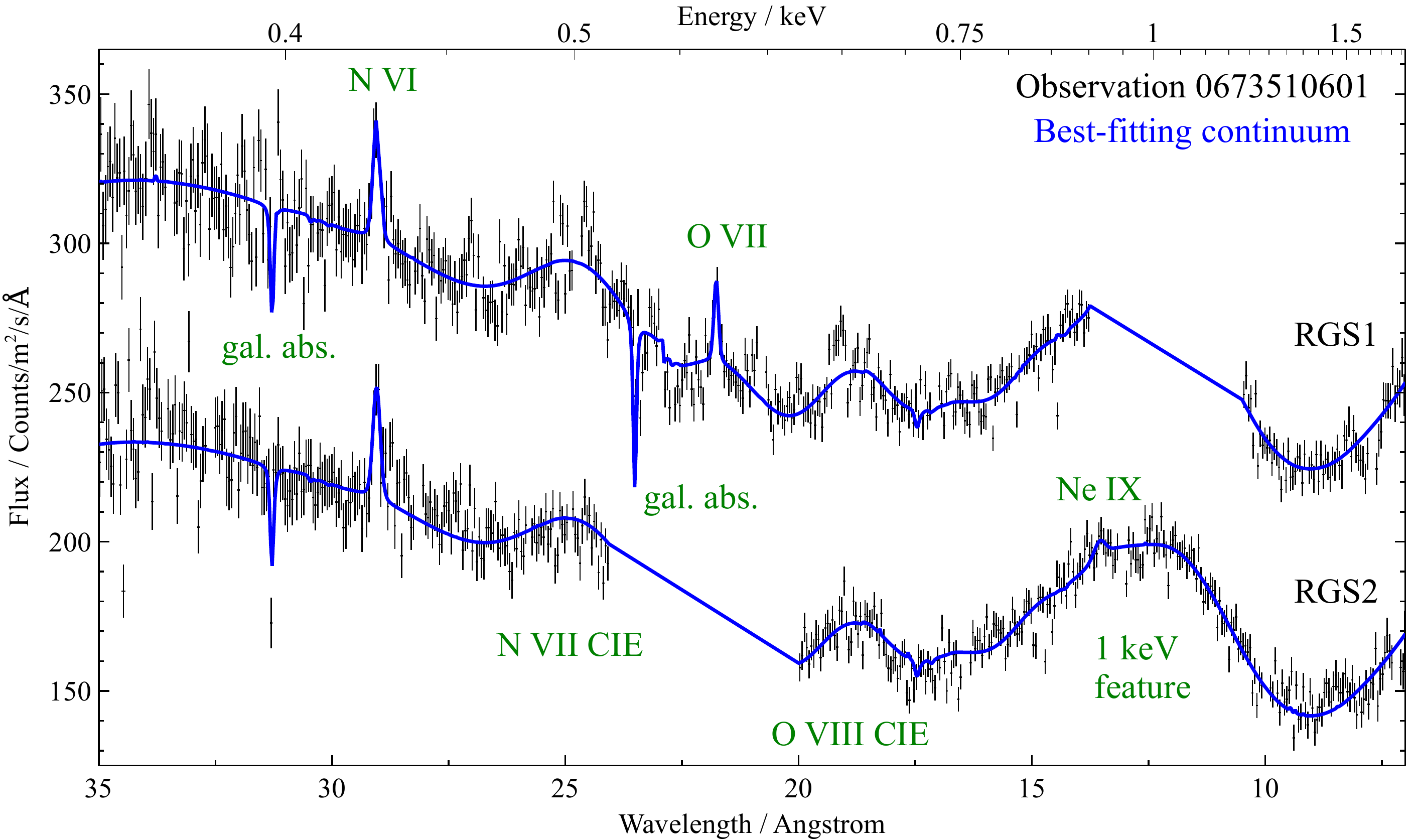}
    \caption{Example 35 \AA\ (0.35 keV) to to 7 \AA\ (1.8 keV) spectrum of Her X-1 using RGS 1 and RGS 2 (Obs ID 0673510601). RGS1 data are shifted by a constant amount for plotting purposes and data from both detectors are heavily overbinned for clarity. Individual spectral components are named with  green labels. Data between 11 and 14 \AA\ in RGS 1 and between 20 and 24 \AA\ in RGS 2 are missing because of chip gaps.}
    \label{Continuum_RGS}
\end{figure*}

First we fit the broadband X-ray spectrum of Her X-1 with an appropriate spectral model. Most of the observations are fitted with the same continuum model, and it will be indicated later where this is not the case. Both RGS and pn data are fitted simultaneously and with the same spectral model within the appropriate energy bands.

A sample pn (3-10 keV) spectrum is shown in Fig. \ref{Continuum_PN}. The high-energy (>3 keV) continuum of the object can be very well reproduced by a Comptonisation model or a powerlaw with an exponential cut-off of about 25 keV \citep{dalFiume+98}. At an even harder band, a cyclotron scattering feature is observed with an energy of about 35-40 keV \citep{Fuerst+13} but this occurs out of the \xmm\ energy band and thus does not need to be taken into account in this study. We describe the broadband continuum with the \textsc{comt} model within SPEX, obtaining a seeding temperature of about $0.05-0.1$ keV, an electron temperature of $\sim3-4$ keV and an opacity of around 10 for all the high state observations.

The Fe K band of the spectrum also contains a strong emission line whose energy and width changes based on the state of Her X-1. In the low state, its energy is $\sim$6.4 keV, with a low ($\sim$0.1 keV) width, whereas in the high state, the line energy is closer to 6.6-6.7 keV with a width of 0.3-0.8 keV \citep{Zane+04}. In this study, for simplicity we fit this region with a Gaussian. One exception is observation 0134120101, where we detect both a narrow (0.1 keV) 6.4 keV emission line and also a broad ($\sim$1 keV) 6.5 keV emission feature - we fit each with a Gaussian.

The physical origin of the broad Fe K line is not understood. The width (up to 0.8 keV), if produced by orbital motion, corresponds to velocities of emitting material of $\sim$0.1c. Such velocities are unlikely to occur in this system given that the accretion disc of Her X-1 is likely truncated by the magnetic field of the neutron star at $\sim$1000 R$_{\textrm{G}}$. \citet{Asami+14} consider different possible origins of the width of the line including line blending, comptonisation from an accretion disc corona or Doppler broadening but do not find a plausible explanation.

The Fe K region in most observations also contains strong residuals which will be well reproduced by highly ionised absorption. Alternatively, they can be fitted with an array of emission lines of iron at various ionisation states.

The soft X-ray (RGS) spectrum is much more complicated. An example observation is shown in Fig. \ref{Continuum_RGS}. A soft ($\sim$0.1 keV) blackbody fits reasonably well the low-energy end of this energy band. However, this is not a signature of the inner accretion disc itself because the spectral component pulses at the pulsar frequency but out of phase compared to the main beam \citep{Endo+00, Ramsay+02, Zane+04}. It is likely that the origin of this component is reprocessed accretion column beam radiation \citep{Hickox+04}. 

Secondly, a strong and broad ($\sim$0.4 keV) Gaussian-like feature is observed at 0.95 keV, also pulsing with the pulsar period. Its origin is currently unknown, but it is suggested to be iron L reflection \citep{Endo+00, Fuerst+13}. We note that the feature is fully resolved in the RGS data and yet its best-fitting spectral model is a simple Gaussian. The feature therefore does not look like an array of unresolved lines, but instead like a single broad feature. Here we describe it with a Gaussian for simplicity. 

Additionally, narrow emission lines can be observed at rest-frame energies of \ion{N}{VI}, \ion{O}{VII} and \ion{Ne}{IX} intercombination lines which suggest the presence of a high density environment. These are especially prominent in the low state of Her X-1 \citep{Jimenez-Garate+02, Jimenez-Garate+05, Ji+09} but still noticeable in the high state.

Adding all of the above components into a continuum spectral fit results into a relatively good fit, however we noticed broad emission residuals at around 19 and 25 \AA\ (Fig. \ref{Continuum_RGS}). Their wavelengths correspond to rest-frame positions of  \ion{N}{VII} and  \ion{O}{VIII} ions. If the residuals were real, they could correspond to photo- or collisionally ionised plasma with large (10000-20000 km/s) velocity widths. Alternatively, these lines could be a signature of blurred reflection. As we do not see residuals of similar strength centred on the rest-frame energies of other N or O lines and/or lines of other elements, it is not possible to distinguish between the first two potential origins of the residuals. Attempting to fit these features with a physical reflection model is beyond the scope of this work. We thus fit the two broad residuals phenomenologically to describe the overall continuum as well as possible. We choose the collisional ionisation emission model \textsc{cie} in SPEX (which is not computationally expensive to fit). We free the normalisation, temperature, velocity width and nitrogen abundance in the model to obtain a simple model with enough freedom, which results in significant fit improvements ($\Delta$C-stat$>$100) in each high state observation. 

The temperature of this plasma  is $\sim$0.25 keV in all observations, with a velocity width of $\sim$15000 km/s and an over-abundance of N/O of about 8-10. Both the velocity width and the N/O ratio seem very high to explain within a system like Her X-1. However, we note that previous studies studies suggest a N/O over-abundance of at least 4 \citep{Jimenez-Garate+05} and that velocities of $10^4$ km/s should not be impossible to achieve at the inner accretion disk/magnetosphere boundary of Her X-1. The orbital velocity at the r=2$\times$10$^{8}~$cm magnetosphere boundary ($\approx$ corotation radius as the neutron star is likely rotating close to equilibrium) of a canonical 1.4 M$_{\odot}$ neutron star is exactly $\sim$10000 km/s. In Fig. \ref{Continuum_RGS} it also appears that the velocity width for both \ion{N}{VII} and \ion{O}{VIII} ions is at least slightly overestimated by our simple model. It is also possible that the N line strength is overestimated compared to the continuum in the 25 \AA\ region. This is the case for multiple high state observations, and could be caused if another spectral component, such as the disc blackbody from the inner accretion disc, is present in the soft X-ray continuum, but is poorly constrained by the present data. If the blackbody temperature is only $\sim$0.05 keV, it would be hard to distinguish given the current energy band and the number of other spectral components present in the soft X-ray spectrum.

The broad emission line component, if real, could therefore originate on the boundary between the inner accretion disk and the magnetosphere of the neutron star. Further studies with future high-spectral resolution instruments like Arcus \citep{Smith+17} should offer sufficient data quality to confirm or reject the presence of these lines and show their origin.

All of the continuum components mentioned above are obscured by interstellar absorption, which we describe with a \textsc{hot} model in SPEX. We set a lower limit of 1.7$\times$10$^{20}~$cm$^{-2}$ to the column density of interstellar gas and fix its temperature to 0.5 eV (neutral gas). The column density value was obtained from  \citet{Kalberla+05}. Finally, we add normalisation constants to RGS 2 and pn datasets to account for calibration differences between the three detectors. Their values are usually very close to 1 (in the 0.95-1.05 range). The final spectral continuum model in SPEX is thus in form of \textsc{hot$\times$(comt+bb+5$\times$gauss+cie)}.

\subsection{Photoionised wind modelling}
\label{high_state}

\begin{table*}
\centering
\caption{Best-fitting wind parameters for each observation of Her X-1. The first column contains the observational ID. The second column lists the measured $0.3-10$ keV unabsorbed luminosity (accounting for the ionised wind absorption). The third column lists the extrapolated $1-1000$ Ryd ($0.0136-13.6$ keV) luminosity, and the fourth the extrapolated $0.0136-80$ keV luminosity. The remaining columns show the properties of outflowing wind such as its column density, ionisation parameter, turbulent velocity and systematic velocity, as well as the statistical fit improvement of the final model compared to the baseline continuum spectral model. (*) in observation 0783770701, we fixed the turbulent velocity of ionised gas to 150 km/s as otherwise it runs away to values much larger than observed in other observations. }
\label{Results}
\setstretch{1.4}
\begin{tabular}{ccccccccc} 
Obs. ID&Luminosity&Luminosity&Luminosity &Column&log $\xi$&Turbulent &Outflow &$\Delta$C-stat\\
&0.3-10 keV&1-1000 Ryd&0.0136-80 keV&density&&velocity&velocity& \\
	& erg$~$s$^{-1}$	& erg$~$s$^{-1}$	& erg$~$s$^{-1}$	&	10$^{24}~$cm$^{-2}$	&	erg$~$cm$~$s$^{-1}$		&	km$~$s$^{-1}$	&		km$~$s$^{-1}$& \\
	\hline
0134120101 &$1.61_{-0.13}^{+0.07}\times10^{37}$&$2.39_{-0.20}^{+0.10}\times10^{37}$&$3.17_{-0.26}^{+0.13}\times10^{37}$&	0.950$_{-0.126}^{+0.016}$&	4.71$_{-0.16}^{+0.21}$&	31$_{-13}^{+26}$&		-270$_{-180}^{+80}$&		10.96\\
0153950301 &$1.25_{-0.11}^{+0.28}\times10^{37}$&$1.87_{-0.16}^{+0.42}\times10^{37}$&$2.50_{-0.22}^{+0.56}\times10^{37}$&	0.17$_{-0.09}^{+0.03}$&	3.96$_{-0.12}^{+0.18}$&	60$_{-30}^{+50}$&		-230$_{-170}^{+150}$&		26.00\\
0673510501 &$9.87_{-0.54}^{+0.68}\times10^{36}$&$1.51_{-0.08}^{+0.10}\times10^{37}$&$2.08_{-0.12}^{+0.14}\times10^{37}$&	0.15$_{-0.07}^{+0.07}$&	3.81$_{-0.08}^{+0.08}$&	140$_{-30}^{+50}$	&	-1000$_{-120}^{+110}$&		88.20\\
0673510601 &$1.58_{-0.14}^{+0.07}\times10^{37}$&$2.43_{-0.21}^{+0.10}\times10^{37}$&$3.61_{-0.31}^{+0.15}\times10^{37}$&	0.35$_{-0.02}^{+0.03}$& 	4.60$_{-0.06}^{+0.07}$&	130$_{-60}^{+100}$ &		-360$_{-110}^{+110}$&		34.69\\
0673510801 &$1.41_{-0.01}^{+0.01}\times10^{37}$&$2.16_{-0.01}^{+0.01}\times10^{37}$&$3.34_{-0.01}^{+0.01}\times10^{37}$&0.006$_{-0.005}^{+0.118}$&3.60$_{-0.17}^{+0.28}$&80$_{-80}^{+510}$&-700$_{-320}^{+400}$&2.72\\
0673510901 &$9.89_{-0.07}^{+0.11}\times10^{36}$&$1.50_{-0.01}^{+0.02}\times10^{37}$&$2.18_{-0.02}^{+0.03}\times10^{37}$&	0.032$_{-0.010}^{+0.006}$&	3.56$_{-0.08}^{+0.07}$&	120$_{-40}^{+90}$	&	-600$_{-100}^{+120}$	&	37.20\\
0783770501 &$5.15_{-0.09}^{+0.09}\times10^{36}$&$7.45_{-0.13}^{+0.13}\times10^{36}$&$9.95_{-0.17}^{+0.17}\times10^{36}$&	0.084$_{-0.021}^{+0.024}$&	2.97$_{-0.05}^{+0.05}$&	75$_{-19}^{+14}$&		-550$_{-180}^{+100}$&		49.86\\		
0783770601 &$8.43_{-0.17}^{+0.18}\times10^{36}$&$1.30_{-0.03}^{+0.03}\times10^{37}$&$1.89_{-0.04}^{+0.04}\times10^{37}$&	0.13$_{-0.05}^{+0.04}$&	3.25$_{-0.07}^{+0.09}$&	70$_{-12}^{+20}$	&	-450$_{-60}^{+110}$&		81.67\\
0783770701 &$1.03_{-0.01}^{+0.03}\times10^{37}$&$1.55_{-0.02}^{+0.04}\times10^{37}$&$2.28_{-0.03}^{+0.06}\times10^{37}$&0.023$_{-0.015}^{+0.074}$&3.67$_{-0.14}^{+0.10}$&150*&-460$_{ -230}^{+190}$&15.13\\
\hline
\end{tabular}
	\\
\end{table*}

In this subsection we model the wind absorption features, measure its physical properties and describe how significant is the wind detection in the X-ray spectrum of Her X-1.

The spectral model used to describe the blueshifted absorption in this section is called \textsc{pion} \citep[see][for more information about the model and its applications]{Miller+15a, Mehdipour+16} in the SPEX fitting package. \textsc{pion} is a powerful photoionisation code that uses the current spectral energy distribution (SED) of all components in the spectral model to calculate the ionisation balance on the fly and can reproduce a broad range of physical parameters of the absorber. Alternatively, the photoionisation absorption model \textsc{xabs} \citep{Steenbrugge+03} could be used to fit the features. It works in a similar way to \textsc{pion} but uses a predefined (active galactic nucleus-like) spectral energy distribution (SED) shape.

We repeat the same process for each observation. Initially, the spectra are fitted with the continuum model described in Section \ref{continuummodelling}. The model parameters as well as the final C-stat value defining the `goodness' of the fit are recovered. Then the \textsc{pion} component is added to the spectral model with appropriate initial parameters. We fit for column density $N_{\textrm{H}}$, ionisation parameter $\log \xi$, turbulent velocity $v$ and systematic (outflow) velocity $z$ of the photoionised absorber. In this section we assume Solar abundances for simplicity. Afterwards, the best-fitting wind parameters as well as the fit improvement $\Delta$C-stat over the original continuum spectral model are recovered.

The best-fitting wind parameters for each observation in our study are listed in Table \ref{Results}. Our results show that the wind velocity varies significantly between the individual observations in the range of 200-1000 km/s. We also find that the ionisation level of the outflowing gas is high with ionisation parameters, $\log \xi$, of 3.0 to almost 5.0. The column density also varies alongside with the ionisation parameter. 

The velocity width of the ionised absorber (from the absorption line widths) is of order of 100 km/s in most observations, with the exception 0783770701, where if freed, it runs away to thousands of km/s (likely due to lack of statistics). The width could be introduced by internal turbulent motion within the flow. Alternatively, it could originate if our line of sight intercepts multiple layers of the wind with a gradient in line of sight velocity over a range of radii. In each case, the velocity width ($\sim100$ km/s) is generally small compared to the line of sight velocity (median value of $\sim$450 km/s), suggesting that the turbulence within the wind is not very strong and that the velocity gradient of all wind layers along the line of sight is not large either.

\begin{figure*}
	\includegraphics[width=\textwidth]{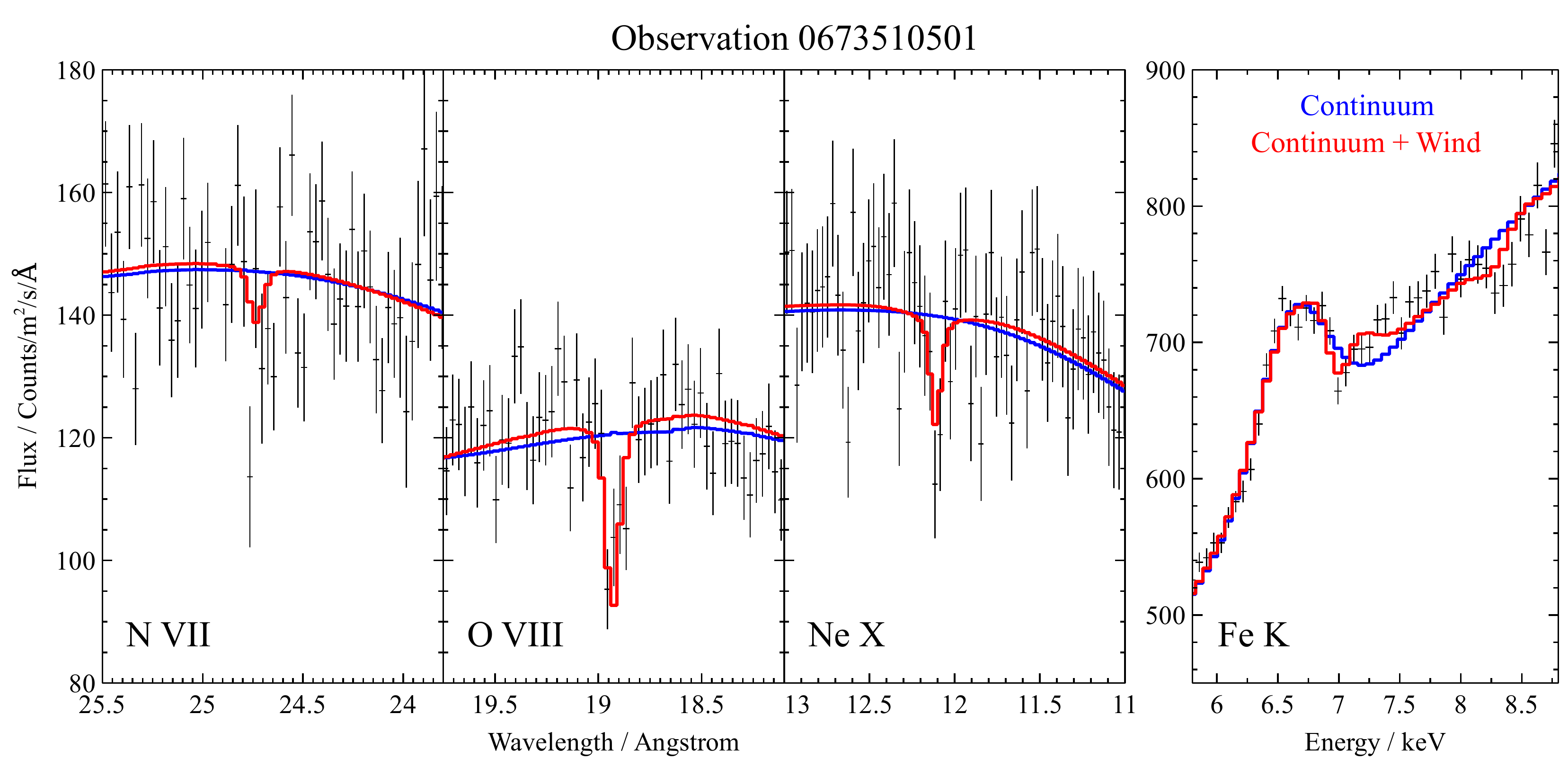}
    \caption{Energy bands around the rest-frame energies of \ion{N}{VII}, \ion{O}{VIII}, \ion{Ne}{X} and \ion{Fe}{XXV/XXVI} ions from observation 0673510501. The first three bands only contain RGS1 and RGS2 data, stacked for plotting purposes only, the fourth band only contains EPIC-pn data. The best-fitting baseline continuum is shown in blue, the final wind solution in red.}
    \label{Spectrum_0673510501}
\end{figure*}

\begin{figure*}
	\includegraphics[width=\textwidth]{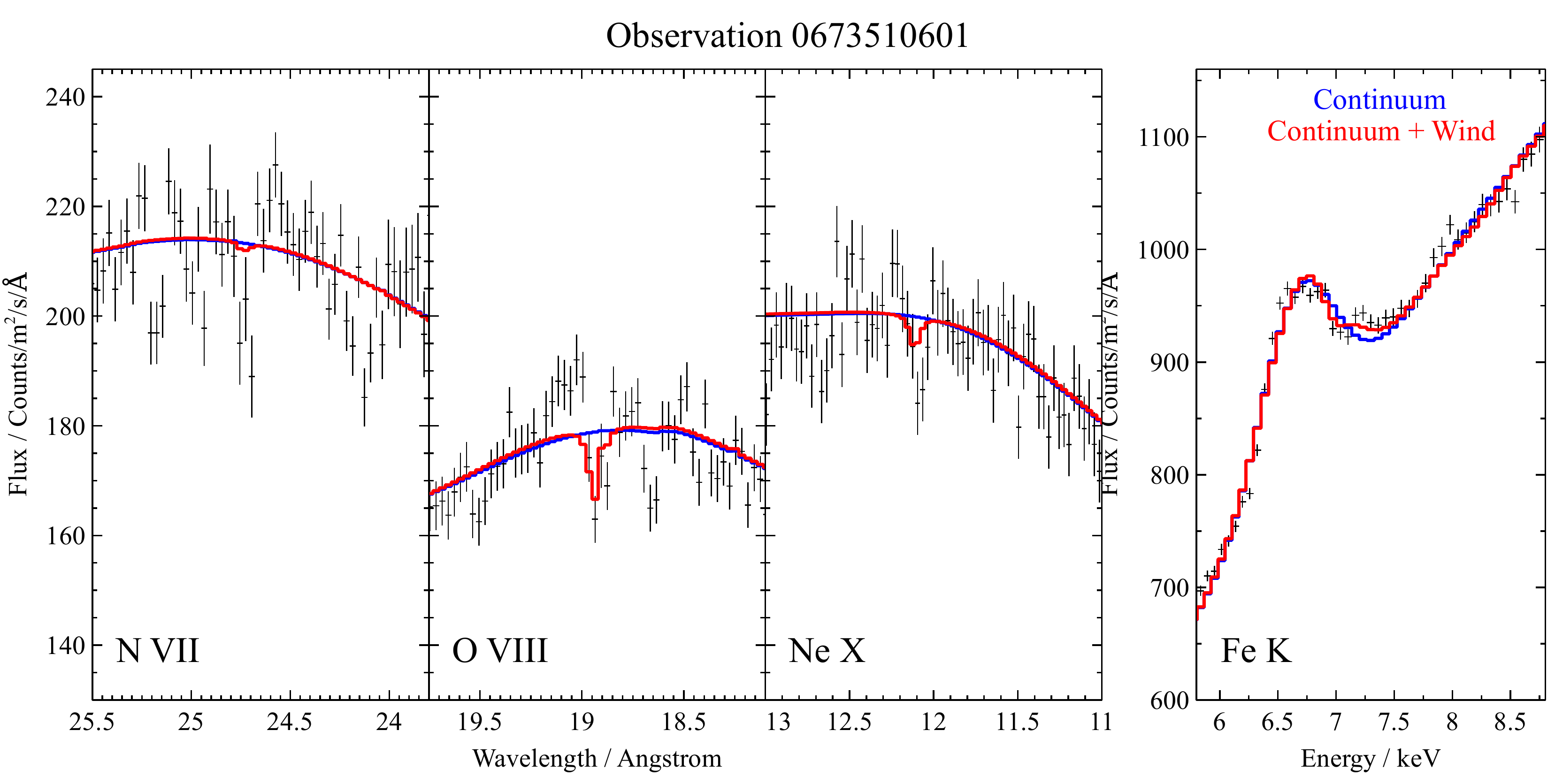}
    \caption{Same plot as Fig. \ref{Spectrum_0673510501} but for observation 0673510601.}
    \label{Spectrum_0673510601}
\end{figure*}

\begin{figure*}
	\includegraphics[width=\textwidth]{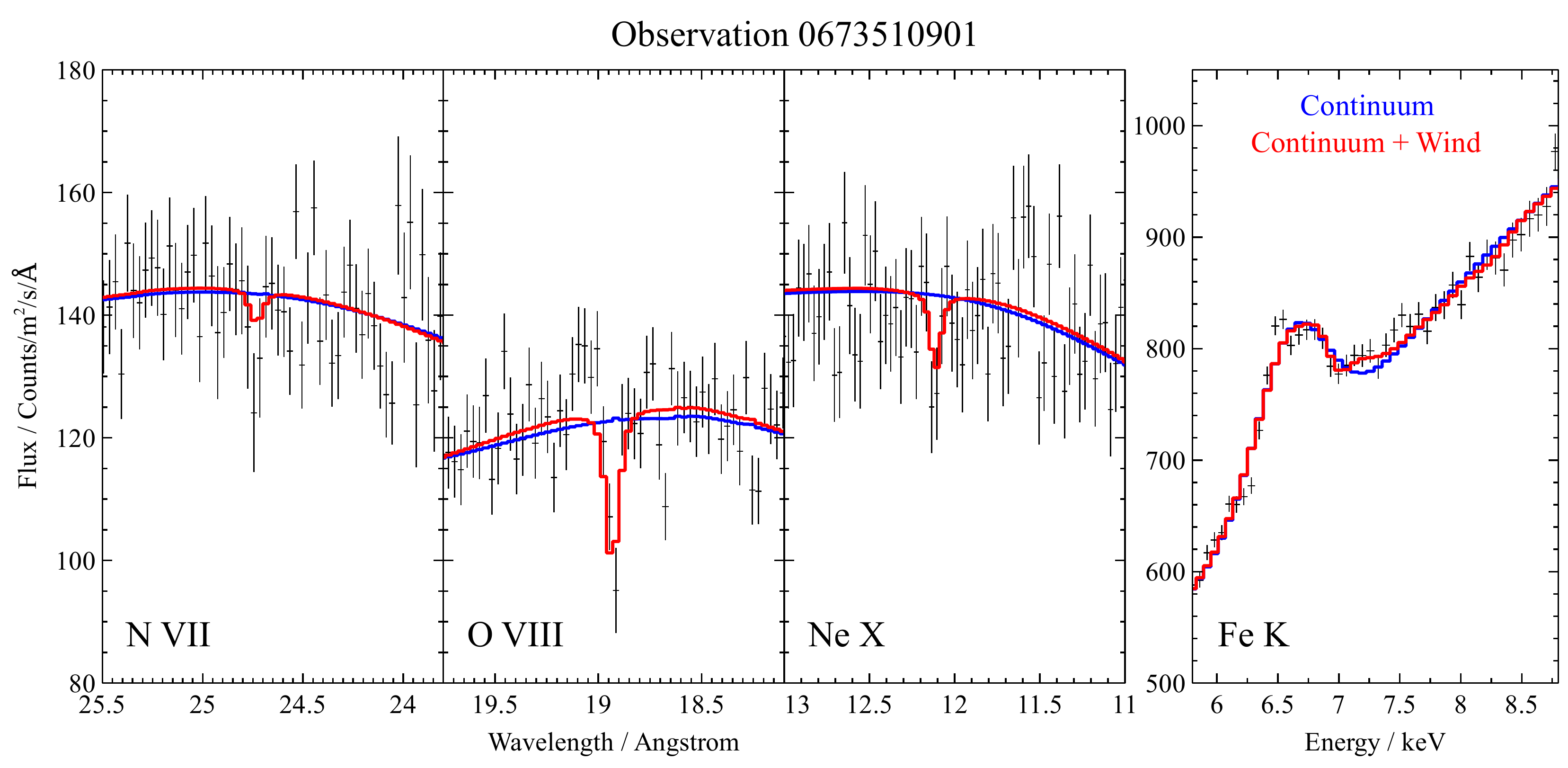}
    \caption{Same plot as Fig. \ref{Spectrum_0673510501} but for observation 0673510901.}
    \label{Spectrum_0673510901}
\end{figure*}

\begin{figure*}
	\includegraphics[width=\textwidth]{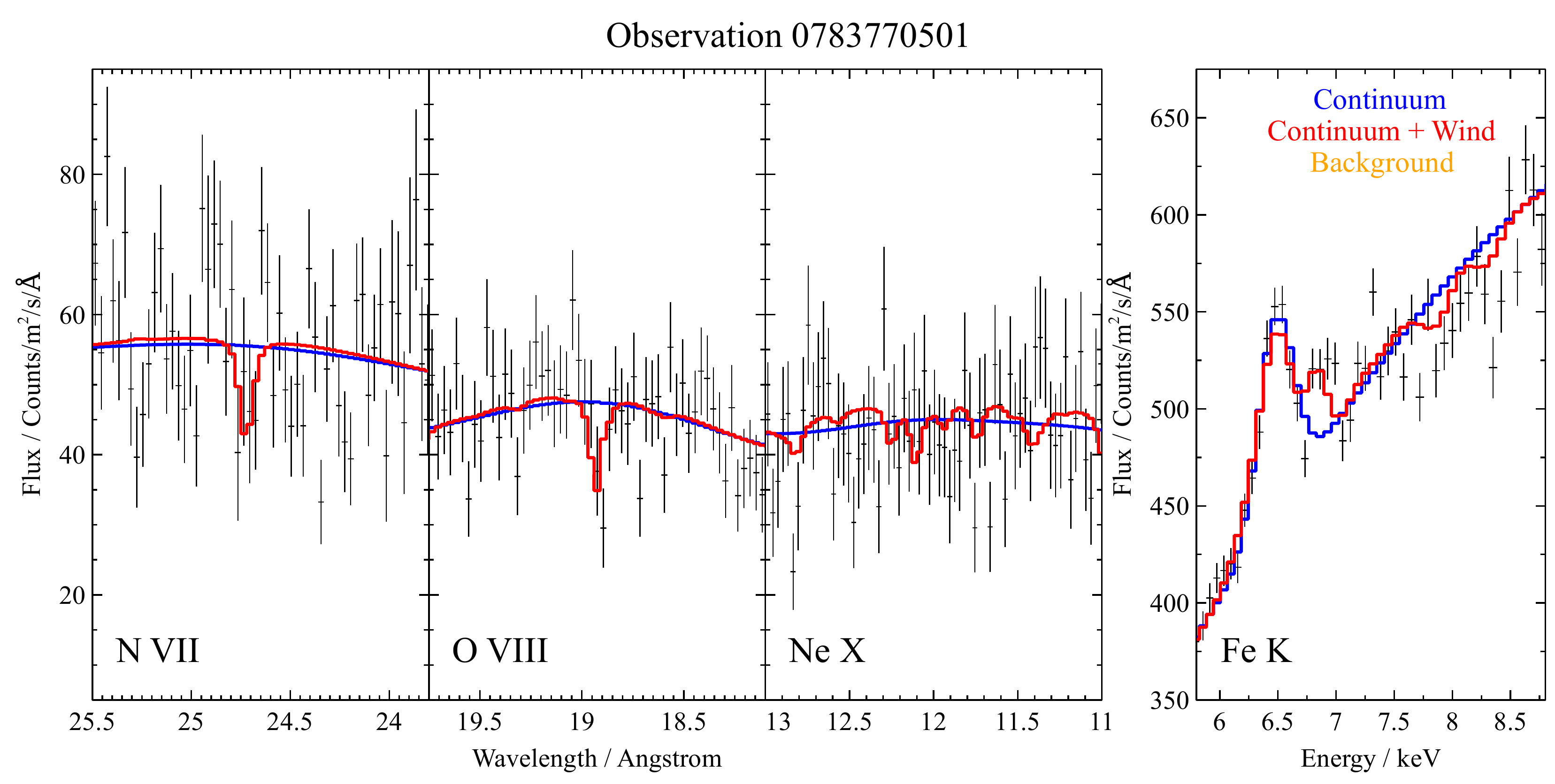}
    \caption{Same plot as Fig. \ref{Spectrum_0673510501} but for observation 0783770501.}
    \label{Spectrum_0783770501}
\end{figure*}

\begin{figure*}
	\includegraphics[width=\textwidth]{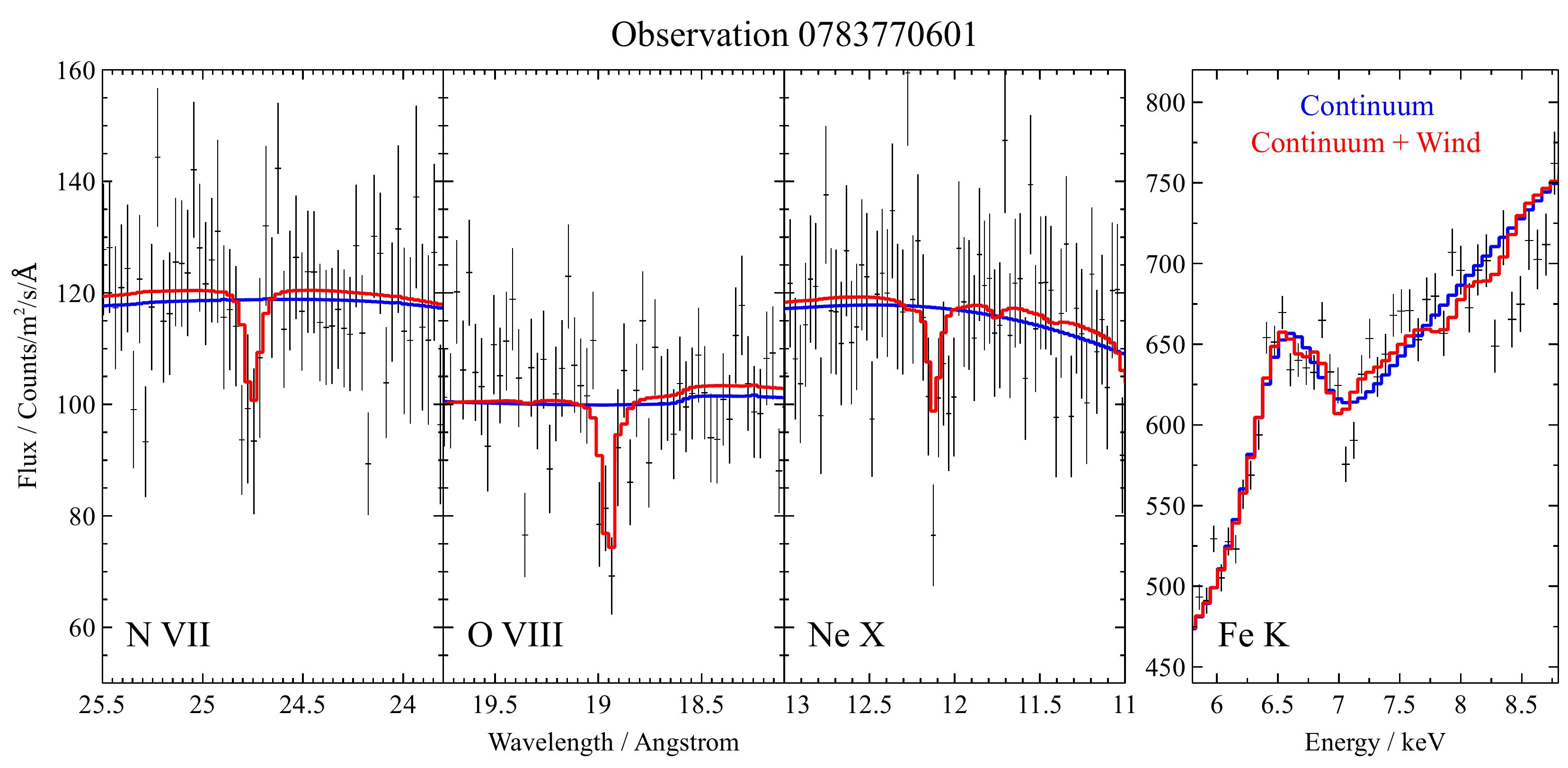}
    \caption{Same plot as Fig. \ref{Spectrum_0673510501} but for observation 0783770601.}
    \label{Spectrum_0783770601}
\end{figure*}

The five observations with the strongest wind features are shown in figures \ref{Spectrum_0673510501} to \ref{Spectrum_0783770601}. At such high ionisation level of the material, the observable features of this wind in our energy band (and considering the CCD resolution of the pn instrument) are only a few high ionisation lines, hence we only show plots containing narrow energy bands around \ion{N}{VII}, \ion{O}{VIII}, \ion{Ne}{X} and \ion{Fe}{XXV/XXVI} line energies. Other strong features of plasma at these ionisation levels are the absorption lines of \ion{Mg}{XII}, \ion{Si}{XIV} and \ion{S}{XVI}, which are occasionally observed in other neutron star \citep[GX 13+1,][]{Ueda+04} and black hole \citep[GRO 1655-40,][]{Miller+06} binaries with similar winds. However, \ion{Si}{XIV} and \ion{S}{XVI} are located in the 2-3 keV energy band which is ignored in this study due to instrumental features in the pn data (RGS band only reaches to 7\AA$\sim$1.8 keV). \ion{Mg}{XII} (at 8.4 \AA$\sim$1.5 keV) is within the RGS band, but at a wavelength where both the spectral resolution and the effective area of the instrument begin to drop. Consequently, the \ion{Mg}{XII} line is not strongly detected.

Unless the abundances of these elements are significantly lower than expected \citep[which has been observed in GRO J1655-40,][]{Kallman+09}, \ion{Mg}{XII}, \ion{Si}{XIV} and \ion{S}{XVI} absorption lines should be detectable with the Chandra HETG gratings, offering a broader (0.3-10 keV) energy bandpass than RGS. Analysis of archival Chandra observations of the high state of Her X-1 will be addressed in our future work.

The statistical fit improvements $\Delta$C-stat vary by a large amount between the individual observations but we consider most detections statistically significant. The strongest detections were achieved in observations 0673510501 and 0783770601, in both cases the $\Delta$C-stat values are $\sim$80. On the other end, the wind was practically undetected in 0673510801 with $\Delta$C-stat=2.7. Other observations with weak detections were 0134120101 and 0783770701.

The statistical significance of the detection of an additional spectral component, in this case of blueshifted absorption can be inferred from the fit improvement $\Delta$C-stat between the two fits (continuum vs continuum + wind). The crucial parameter here is the number of additional free parameters that the wind spectral models adds (in our case this is 4 - column density, ionisation parameter, turbulent and systematic velocity of the absorber). However, since the continuum model is effectively on the edge of the parameter space of the more complicated, continuum+wind model (where the column density of the ionised absorber is simply 0), it is not possible to determine the significance rigorously by a theoretical approach like an F-test \citep{Protassov+02}. The solution is to perform Monte Carlo simulations -  first a blind search is ran over the wind parameters using the real data. Then a similar dataset containing only the continuum model spectrum is simulated, and the same wind search is ran on the simulated data. The statistical significance of the detection of a wind in the real dataset is then 1 minus the fraction of occurences of detections in fake data stronger (with larger $\Delta$C-stat values) than the $\Delta$C-stat of the real detection.

It is not computationally feasible to perform such a search in this situation and assess the detection significance completely rigorously. This is because the underlying spectral continuum of Her X-1 is too complex, causing the simulated blind search to become very computationally expensive. However, we would like to compare the fit improvements seen in this study with the results from \citet{Kosec+18b}, where a full Monte Carlo simulation suite was performed. In that study a wind with $\Delta$C-stat of $\sim$27 was detected using 4 additional free wind parameters (the same number as here). However, they used a much wider parameter space - systematic velocity space of 0 km/s to 120000 km/s, whereas in this study we only assess wind velocities from 0 to a few thousands km/s. The statistical significance of wind detection in \citet{Kosec+18b} was about 3.5$\sigma$. We therefore argue that the wind detection in most of the observations of Her X-1 is statistically significant.

Photon pile-up could affect some of our datasets, especially pn and RGS2. We address this issue in Appendix \ref{Appendix}. However, it is unlikely that it could introduce absorption features which line up in the velocity space. We conclude that our detection of an ionised wind in the spectrum of Her X-1 is robust.

We also consider a possibility that the outflowing plasma is multiphase, i.e. it has multiple ionisation and velocity components. This seems to be the case for other X-ray binaries with wind detections \citep[e. g.][]{Miller+15b}. We can exclude significant absorption by low ionisation ($\log \xi<2$) material, which would have a strong signature in the soft X-ray (RGS) band. Unfortunately, most of our observations do not offer high enough data quality (statistics) to address this hypothesis for the higher ionisation levels. This is due to low column densities of the wind given the ionisation state and consequently low optical depths of absorption features (e. g. Fig. \ref{Spectrum_0673510901}). Additionally, at such high ionisation levels ($\log \xi=3-5$), there is only a small number of strong lines left in the absorption spectrum. In most observations we thus do not have enough photon counts to distinguish multiple wind components, despite the high flux of Her X-1. Future longer exposure observations may address this problem.

We attempt to test the multiphase hypothesis at least in the observations where the wind absorption is the strongest. Choosing the two observations with the highest \delcstat\ fit improvement upon adding the ionised absorption component (0673510501 and 0783770601), we fit the spectra with a double ionised absorption model. We use the \textsc{pion} spectral model to describe the two absorption zones, with all relevant physical parameters decoupled. In the case of observation 0673510501, we find a very small fit improvement of $\Delta$C-stat$\sim$6 compared to the single zone absorption model. In the case of observation 0783770601, the fit improvement is $\Delta$C-stat$\sim$14, larger but still not statistically significant to warrant the addition of a second absorption zone. In conclusion, with the current data quality there is no strong evidence for a multiphase nature of the wind in Her X-1.

\subsection{Short-on and low state observations}

We checked the low and short-on state observations of Her X-1 for any obvious signatures of blueshifted absorption lines. Naturally, the flux and hence the count rate during these epochs are much lower than in the high state. None of the observations individually can be used to constrain the presence of the wind - the continuum is too weak in the RGS band and only photoionised line emission is detected significantly. We omit the Fe K band as it is more complicated than in the high state, with a 6.4 keV narrow line (\ion{Fe}{I}), a possible 6.97 keV line (\ion{Fe}{XXVI}) plus an edge at 7.1 keV due to a partial covering absorber \citep{Ji+09}.

To improve the statistics, we stack all the available low state RGS data for observations which are not affected too strongly by background flares, for a clean exposure of roughly 45 ks per detector. This allows us to get a significant detection of the X-ray continuum, nevertheless we do not observe any obvious absorption features at similar systematic velocities ($0-1000$ km/s) compared to the ones seen in the high state. The search is naturally complicated by the fact that the stack comes from averaging over several years of observations. If the wind velocity is time variable, even if the absorption is present in the low state, its features would be smeared and difficult to detect using the stacked dataset. The analysis is further complicated by the strong low state line emission, which is challenging to model physically. Any possible absorption features will be difficult to disentangle from the emission lines which are at similar energies (outflow velocity of 1000 km/s corresponds to a wavelength shift of just $3\times10^{-3}$).  A rigorous search of the low state data is beyond the scope of this work.

\section{Elemental Abundances in the Wind}

So far we have assumed solar abundances while fitting the wind parameters. However, the outflowing material can serve as a powerful probe to independently determine the composition of the matter accreted onto Her X-1. Previous studies based on line emission in the low state suggested over-abundance of N and Ne compared to O \citep{Jimenez-Garate+05}. 

The measurement of abundances from blueshifted absorption lines is in principle a much easier task, however given the high ionisation level of the wind, the absorption spectrum only contains a few metallic lines and there is very little continuum absorption. This has two important consequences given the current data quality. 

First, we choose to perform a simultaneous fit of multiple observations at once to increase the signal-to-noise. We free all of the individual observation model parameters with the exception of wind material abundances (within the \textsc{pion} model) and simultaneously fit 5 observations with the strongest wind detection - 0673510501, 0673510601, 0673510901, 0783770501 and 0783770601. We could in principle fit all of the available observations simultaneously, but this would be too computationally expensive.

The second important consequence is that this analysis is unable to measure absolute abundances. It can only constrain relative abundances of elements compared to one selected element whose absorption line is strong enough to anchor the spectral fit. The only elements with strong enough lines present in photoionised spectra at this ionisation level and in our energy band are N, O, Ne and Fe (O and Fe being the strongest). We therefore follow two avenues: first, we fix the iron abundance and measure relative abundances of N, O and Ne compared to Fe; afterwards we fix the oxygen abundance and fit for N, Ne and Fe. Ideally, these two approaches should result in similar elemental ratios and serve as an independent check.

It is not obvious how to treat the abundances of the remaining elements such as Mg, Si, S, Ni and others. At Solar abundances, their absorption lines are weak. However, once we free the abundances of the main elements, they might become important. Initially, we freeze their abundance to 1. This effectively means that the abundances of these elements are equal to that of the comparison element (Fe or O).

First, we freeze the abundance of Fe. We recover an over-abundance of N, Ne and Fe of 2 to 4 compared to O, for a modest fit improvement of $\Delta\textrm{C-stat}\approx20$ (1st row of Table \ref{Results_abundances}). Secondly, we freeze the abundance of O. In this case we find a much larger fit improvement of $\Delta\textrm{C-stat}\approx120$ and also much higher elemental ratios (3rd row of Table \ref{Results_abundances}). The Fe/O ratio is the highest at $17.1_{-1.2}^{+1.5}$.  We interpret this significant difference between the fit quality in these two approaches to be caused by the remaining elements whose abundances are frozen. When O is freed, its abundance begins to decrease compared to Fe. However, the \ion{O}{VIII} line is the strongest wind absorption line, so to remain fitted correctly, the column density of the wind material must be increased, thus strengthening the absorption lines of all the frozen elements (whose lines are weak in the actual spectrum). We conclude that this suggests that the abundance O is not under-abundant compared to these elements, and hence this approach to fitting the abundances is not trustworthy.

We also experiment with setting the abundances of the `weak' elements to 0. While this is an unphysical scenario, it approximates a situation in which the abundance of the comparison element (which is frozen to 1) is much larger than the abundance of the `weak' elements (without adding too much computational cost). We find similar results regardless of whether Fe or O is the comparison element (2nd and 4th rows of Table \ref{Results_abundances}), as is expected. N and Ne appear to be over-abundant compared to O by a factor of 2 to 4, and the Fe/O ratio is as high as 10.

\begin{table*}
\centering
\caption{Best-fitting abundances of elements and elemental ratios for each of the four approaches to the chemical analysis. The last column contains the fit improvement in $\Delta$C-stat for each approach.}
\label{Results_abundances}
\setstretch{1.4}
\begin{tabular}{ccccccccc} 
N&O&Ne&Fe&Other&N/O&Ne/O&Fe/O&$\Delta$C-stat\\
&&&&elements\\
\hline
$1.6_{-0.6}^{+0.7}$&$0.44_{-0.08}^{+0.06}$&$0.96_{-0.17}^{+0.22}$&1*&1*&$3.6_{-1.4}^{+1.7}$&$2.2_{-0.5}^{+0.7}$&$2.3_{-0.3}^{+0.4}$&19.74\\ 
$0.21_{-0.07}^{+0.08}$&$0.086_{-0.018}^{+0.024}$&$0.36_{-0.13}^{+0.09}$&1*&0*&$2.4_{-1.1}^{+1.1}$&$4.2_{-1.9}^{+1.4}$&$11.6
_{-3.3}^{+2.4}$&121.16\\
$3.9_{-1.3}^{+1.6}$&1*&$5.6_{-1.7}^{+2.0}$&$17.1_{-1.2}^{+1.5}$&1*&$3.9_{-1.3}^{+1.6}$&$5.6_{-1.7}^{+2.0}$&$17.1_{-1.2}^{+1.5}$&118.40\\ 
$1.9_{-0.6}^{+1.0}$&1*&$2.5_{-0.7}^{+1.2}$&$9.1_{-0.9}^{+1.3}$&0*&$1.9_{-0.6}^{+1.0}$&$2.5_{-0.7}^{+1.2}$&$9.1_{-0.9}^{+1.3}$&134.03\\ 
\hline
\end{tabular}
\\
* The elemental abundance is fixed to the corresponding value during the fit.
	\\
\end{table*}

In conclusion, our results confirm the previous findings of \citet{Jimenez-Garate+02, Jimenez-Garate+05} regarding the elemental ratios of N/O and Ne/O. We find that the N/O ratio is between 2 and 4 for different approaches to the fitting analysis. Ne is also over-abundant compared to O, we find that $\textrm{Ne/O} \approx2-6$, in line with previous results.

Unexpectedly, we also find very high Fe/O ratios. The exact ratio heavily depends on the approach chosen - we obtain $\textrm{Fe/O} \approx2$ for Fe fixed to 1 (but do not trust this result due to reasons given above), $\textrm{Fe/O} \approx10$ for both approaches with the remaining elements fixed to 0, and $\textrm{Fe/O} \approx17$ for O fixed to 1. We suspect that the last value is a strong overestimate, possibly driven by the abundance of the `weak' frozen elements. We prefer the results from the approaches where the `weak' elements are set to 0 and argue that the Fe/O ratio might be as high as $8-10$. This is still very high but probably more realistic than $>15$. We find that these two approaches result in very similar elemental ratios and $\Delta$C-stat fit improvements, as expected because they should be almost equivalent.

We however stress one important point - the Fe abundance measurements at these ionisation levels are all based on the Fe K energy band. Our spectral resolution in this band is modest ($\sim$100 eV resolution of the pn instrument) and its modelling is quite simplistic. If the true underlying spectral model is significantly more complicated than assumed in this work (i.e. if there is a range of Fe emission lines at 6.4 keV, 6.7 keV and 6.97 keV compared to one broad Gaussian line), the Fe/O elemental ratios obtained here must be taken with caution. Finally, we note that we have assumed that the gas is in equilibrium, which might not be entirely true (for example if the wind is driven along magnetic lines).

We conclude that the abundances in Her X-1 are strongly non-Solar. This is evidenced by the large fit improvements ($\Delta\textrm{C-stat} > 100$) upon freeing the abundance parameters.

\section{Discussion}

We have shown that the X-ray spectrum of Her X-1 during the high state contains strong evidence of blueshifted wind absorption. The Fe K band of the spectrum by itself could be explained by an array of Fe emission lines (at 6.4, 6.7, 6.97 keV) rather than by absorption features \citep{Asami+14}. However, the \ion{N}{VII}, \ion{O}{VIII} and \ion{Ne}{X} regions unambiguously show blueshifted absorption lines at the same systematic velocity, thus confirming that we are observing an ionised wind. The wind detection is statistically significant in most of the \xmm\ observations with the exception of 0673510801 and 0134120101, where the evidence for absorption features is weaker. At this moment we do not find evidence for similar blueshifted absorption in the short-on and low states of Her X-1, but the data quality of these observations is much lower. Stacking multiple datasets likely smears the absorption signatures as the wind appears to be variable in time.

We will now investigate how the wind parameters vary across different high state observations. Afterwards, we will estimate the launching radius of the wind as well as the mass outflow rate. Finally, we will attempt to pinpoint its launching mechanism and try to explain the variation of wind parameters in time.

\subsection{Wind evolution with luminosity, orbital and super-orbital phase}

For these calculations, it is necessary to obtain the luminosity of the ionising radiation of the object. The wind naturally sees the full energy band of radiation and not just the luminosity in the RGS and pn band (0.3-10 keV, listed in the $2^{\textrm{nd}}$ column of Table \ref{Results}). By definition the 1-1000 Ryd energy band is taken when considering the ionising flux. For this reason we calculate the extrapolated $0.0136-13.6$ keV luminosity of Her X-1 for each observation. The errors introduced by this extrapolation should not be large on the upper energy end as pn data reaches to 10 keV. At the low energy end, we neglect the extreme UV radiation from the accretion disc whose spectrum does not reach into the X-ray band and thus cannot be constrained by \xmm. However, given that the disc is truncated at $\sim$1000 R$_{\textrm{G}}$, the systematic error introduced by this simplification should not be huge. The ionising luminosity estimates are shown in Table \ref{Results} ($3^{\textrm{rd}}$ column). Finally, we also calculate the total luminosity of Her X-1 for each observation by extrapolating between $0.0136-80$ keV ($4^{\textrm{th}}$ column of Table \ref{Results}). These estimates should be taken with some amount of caution.

\begin{figure}
	\includegraphics[width=\columnwidth]{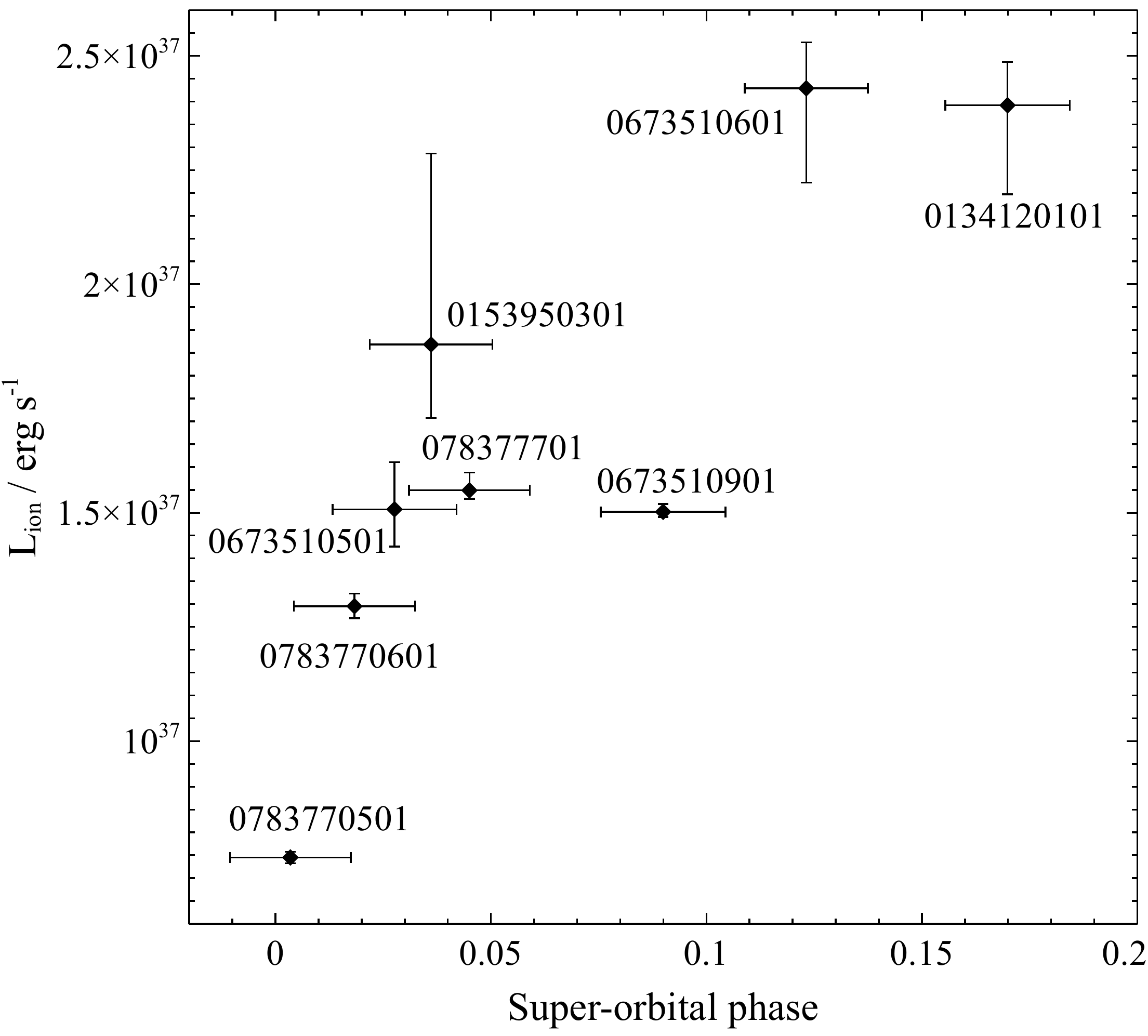}
    \caption{Extrapolated 1-1000 Ryd ionising luminosity for each of the high state observations versus the super-orbital phase. }
    \label{Lion_suporb}
\end{figure}

We plot the ionising luminosity with respect to the super-orbital phase for each observation in Fig. \ref{Lion_suporb}. The range of luminosities sampled by \xmm\ observations nicely reproduces the high state part of the super-orbital flux lightcurve of Her X-1 \citep[e.g. Fig. 2 from][]{Leahy+10}. The only outlier is observation 0673510901, during which the luminosity is around 50\% smaller than would be predicted by fitting all the other data points. To investigate this outlier we checked the pn lightcurve of observation 0673510901 but found no evidence for discrete obscuration events.

We note that in the next part of the study we omit observation 0673510801 results as the wind detection is insignificant and its uncertainties are too large for any informative conclusions.

\begin{figure}
	\includegraphics[width=\columnwidth]{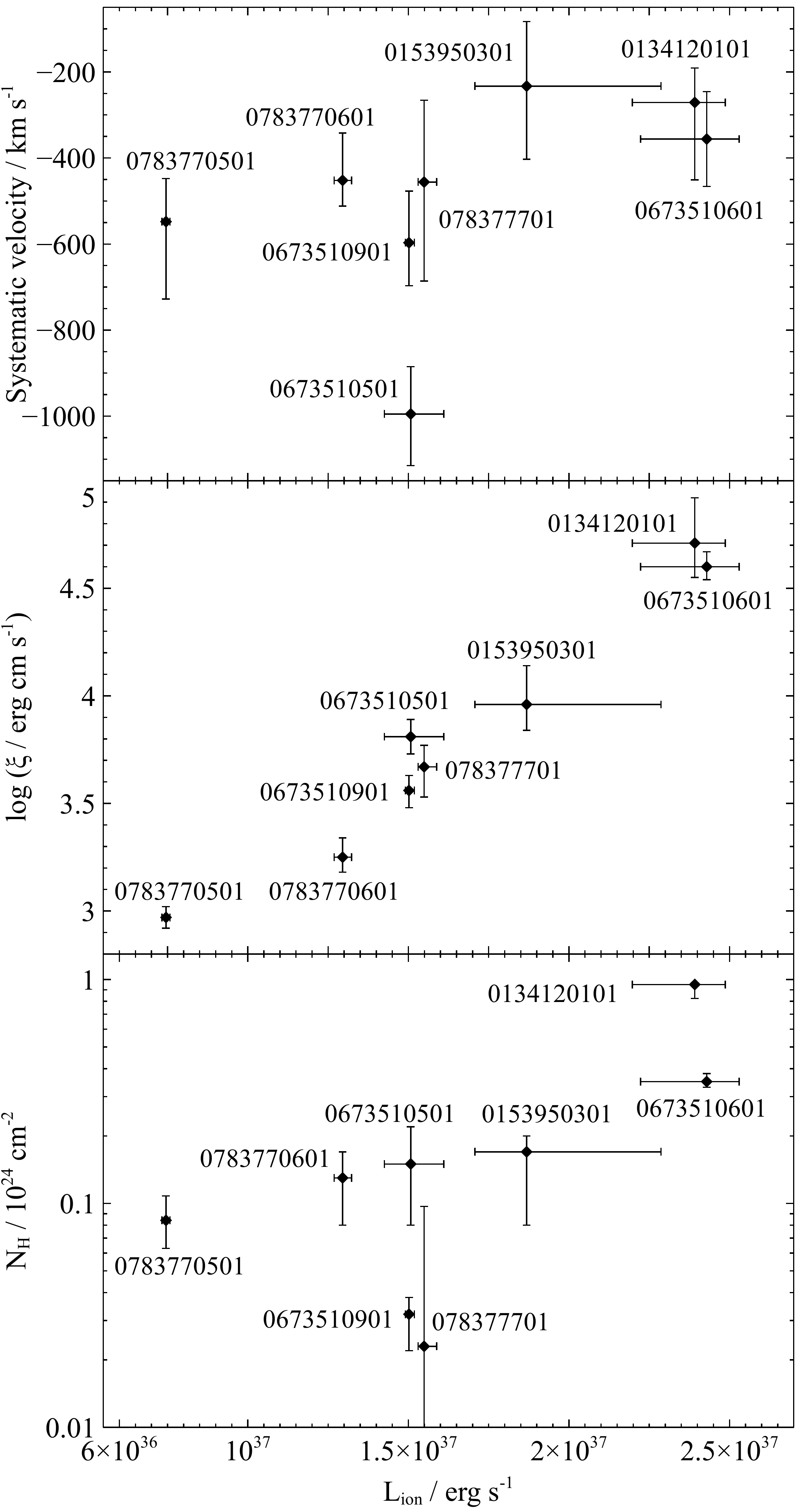}
    \caption{\textit{Top plot:} Systematic velocity of the ionised absorber with respect to the extrapolated 1-1000 Ryd ionising luminosity of Her X-1 for each observation in the high state. \textit{Middle plot:} Ionisation parameter of the absorber versus the extrapolated 1-1000 Ryd luminosity for each observation in the high state. \textit{Bottom plot:} Column density of the absorber versus the 1-1000 Ryd luminosity for each observation.}
    \label{Lion_zv_xi_nh}
\end{figure}

The projected wind velocity spans a range of velocities between 200 and 1000 km/s and is inconsistent with being constant across all the observations. Fig. \ref{Lion_zv_xi_nh} (top plot) shows that there does not seem to be a clear correlation between the outflow velocity and the 1-1000 Ryd luminosity. Observation 0673510501 appears to be an outlier during which the wind was apparently much faster. 

On the other hand, there is a clear positive correlation between the ionisation parameter and the luminosity of Her X-1, shown in Fig. \ref{Lion_zv_xi_nh} (middle plot). Such correlation suggests that the wind responds to the change in luminosity of the object and thus sees similar if not the same luminosity as we observe. This was not a given because the change in Her X-1 luminosity is likely only an obscuration or projection effect and the accretion onto the primary continues at a nearly constant pace.

There is a tentative correlation between the wind column density and the ionising luminosity (Fig. \ref{Lion_zv_xi_nh}, bottom plot), but with clear outliers - observations 0673510901 and 0134120101.

\begin{figure}
	\includegraphics[width=\columnwidth]{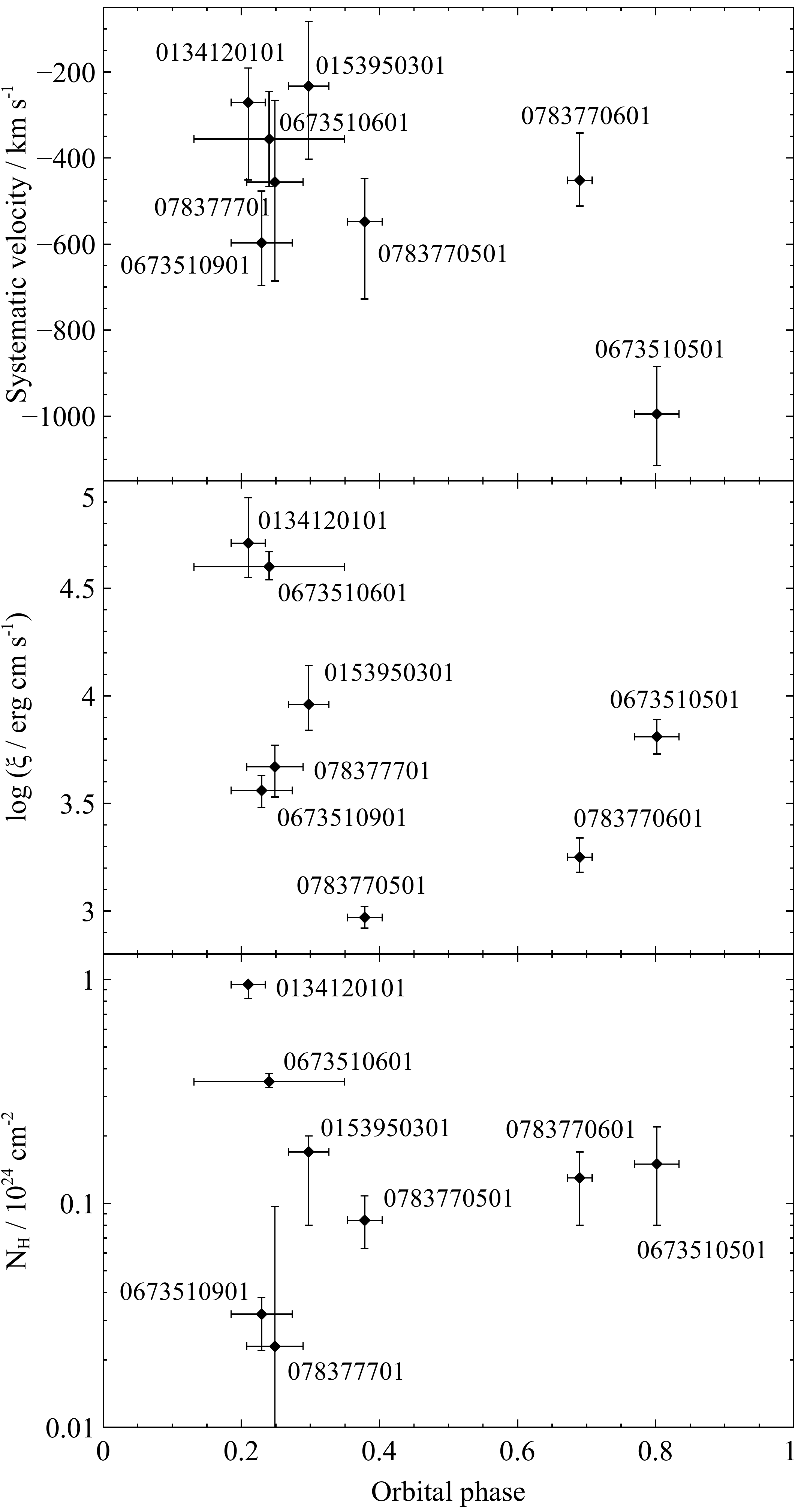}
    \caption{\textit{Top plot:} Systematic velocity of the ionised absorber with respect to the orbital phase during each observation. \textit{Middle plot:} Ionisation parameter of the absorber versus the orbital phase during each observation. \textit{Bottom plot:} Column density of the absorber versus the orbital phase during each observation. }
    \label{Windpar_orb}
\end{figure}

We do not observe any significant correlations between the wind parameters and the orbital phase of each observation. This finding suggests that the wind is not tied in any way to the secondary of the binary system or the motion of the primary and is only related to the accretion disc of the neutron star. The wind parameters with respect to the orbital phase are shown in Fig. \ref{Windpar_orb}.

\begin{figure}
	\includegraphics[width=\columnwidth]{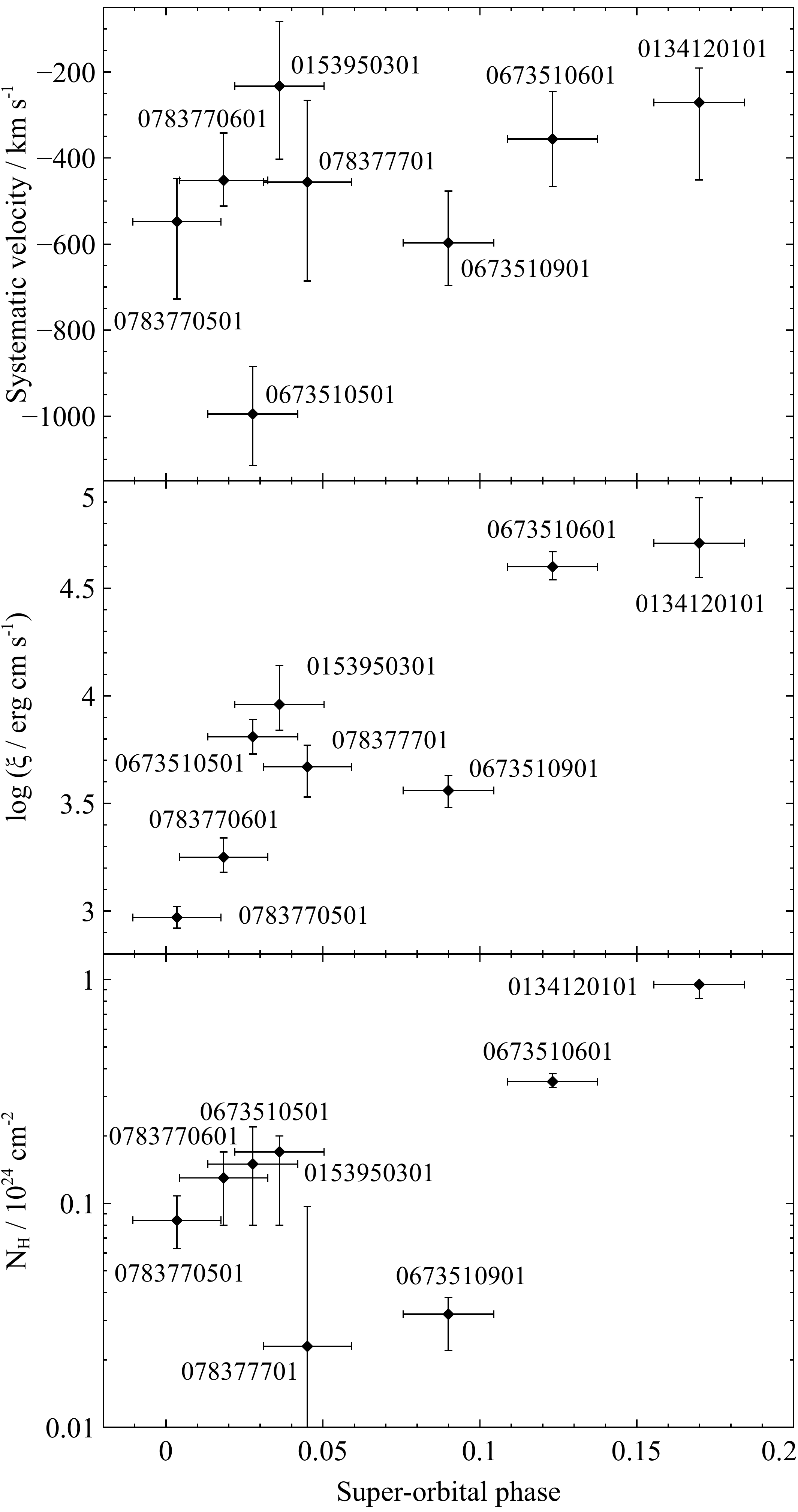}
    \caption{\textit{Top plot:} Systematic velocity of the ionised absorber with respect to the super-orbital phase during each observation. \textit{Middle plot:} Ionisation parameter of the absorber versus the super-orbital phase during each observation. \textit{Bottom plot:} Column density of the absorber versus the super-orbital phase during each observation. }
    \label{Windpar_suporb}
\end{figure}

Fig. \ref{Windpar_suporb} shows the wind parameters versus the super-orbital phase of each exposure. As with the ionising luminosity, we do not find any correlation between the outflow velocity and the super-orbital phase. We notice a strong correlation between the ionisation parameter and the super-orbital phase, with one obvious outlier - observation 0673510901. This correlation comes naturally since the ionising luminosity is correlated with the super-orbital phase and the ionisation parameter is correlated with the 1-1000 Ryd ionising luminosity. However, the fact that observation 0673510901 is an outlier in the super-orbital plot and not in the luminosity plot could suggest that the ionisation parameter depends on the luminosity and not on the super-orbital phase. This could mean that we are not only probing different lines of sight from the neutron star (as the super-orbital phase progresses), but that the ionising flux on the wind gas must also change in time. Alternatively, observation 0673510901 could be an outlier in the super-orbital cycle, an anomalous state. Finally, there is a tentative correlation between the column density of the outflowing material and the super-orbital phase, with two outliers being observations 0673510901 and 078377701.

\subsection{Location of the wind absorption}
\label{Dist_section}

\begin{figure*}
	\includegraphics[width=\textwidth]{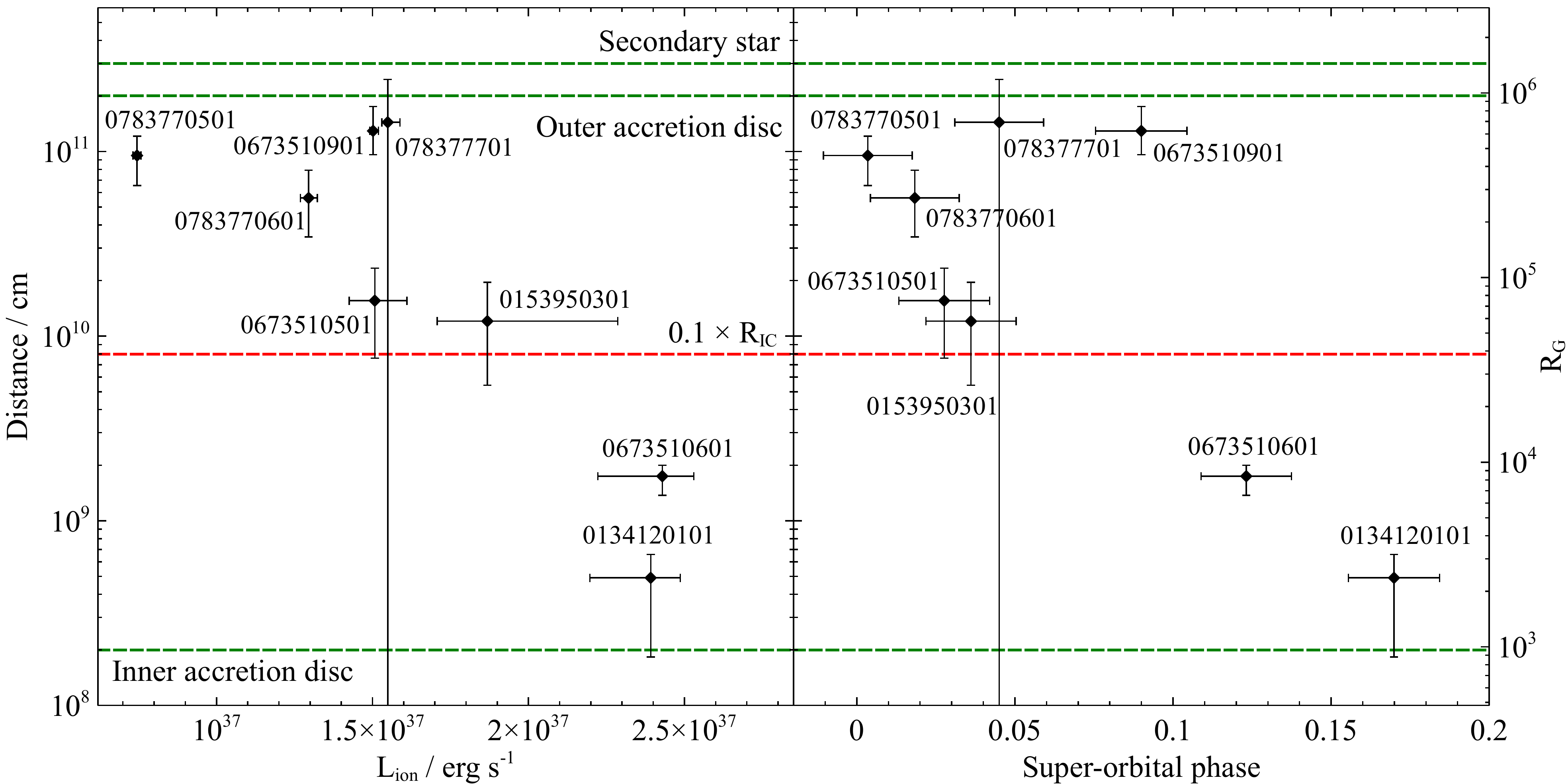}
    \caption{The estimate of the maximum distance of photoionised absorption from the ionising source versus the ionising luminosity (left subplot) and the super-orbital phase (right subplot). The green horizontal lines show the positions of the inner and outer edges of the accretion disc, and the distance of the secondary star from the primary. The red dashed line shows the approximate position of the minimum wind launching radius if the wind is powered by Compton heating of the accretion disc.}
    \label{Distance_Lion}
\end{figure*}

\citet{Boroson+01} found blueshifted absorption UV lines in Her X-1 with a velocity of several hundred km/s using the FOS and STIS spectrographs onboard the Hubble Space Telescope. They concluded that the wind is likely launched from the X-ray irradiated side of the secondary, and it is observed at larger, circumbinary distances from the system. The conclusion was motivated by the stability of the wind systematic velocity over the orbital cycle of the binary.

We similarly attempt to pinpoint the location of the X-ray absorber and its relation to the UV absorber, as well as the likely launching mechanism. The high ionisation level of the gas by itself suggests the proximity of absorption to the ionising (X-ray emitting) source. We can put an upper limit on the distance of the wind absorption from the neutron star by using the ionisation parameter and the column density of the absorber.

We will assume that the wind is in equilibrium with the ionising radiation. The unabsorbed 1-1000 Ryd luminosity is used as the ionising luminosity as opposed to a constant $4 \times 10^{37}$ erg/s luminosity because the ionisation parameter is seen to respond to the luminosity change. The ionisation parameter of the absorber is then defined as:

\begin{equation}
\xi=\frac{L_{ion}}{nR^{2}}
\label{eqxi}
\end{equation}

where $L_{ion}$ is the ionising luminosity, $n$ is the ion number density and $R$ is the distance from the ionising source \citep{Tarter+69}. The column density can be expressed as:

\begin{equation}
N_{\textrm{H}}=n  \Delta R = n  R \delta R
\label{eqnh}.
\end{equation}

where $\Delta R$ is the thickness of the absorbing layer and $\delta R = \Delta R / R$ is its relative thickness ($\delta R<1$). We can hence express $R$ as:

\begin{equation}
R=\frac{L_{ion}}{N_{\textrm{H}}\xi} \delta R
\label{eqdist}
\end{equation}

This is naturally a very simple calculation assuming that the absorbing layer is uniform and gives an approximate upper limit on the distance of the absorber from the neutron star. If the absorption occurs over a layer that is not very thin, the fractional distance $\delta R = 0.1-1$ and thus the $\delta R$ correction factor is not going to be too important. The maximum distance for each Her X-1 observation is shown in Fig. \ref{Distance_Lion} with other relevant scales of the system and is also listed in Table \ref{Results_Dist_Mass}.

All of the results clearly lie within the accretion disc of the primary, and they do not seem to cluster on the inner edge of the disc. A few results (0783770701, 0673510901) are consistent with the outer disc boundary or the secondary star surface, but most are completely inconsistent with this interpretation. We note that even if the correction factor $\delta R$ is mild (0.1-0.5), it will push most of the outlying results down inside the accretion disk. We therefore argue that the absorption occurs within the size of the accretion disk of the primary. The wind must hence be launched from the accretion disc (or even closer to the neutron star) and can not originate on the irradiated side of the secondary. Our interpretation is that the wind originates within the accretion disc with a speed higher than the escape velocity and is thus able to leave the system, imprinting UV absorption features on the spectrum of Her X-1 at larger (circumbinary) distances. 

On the other hand, unless the thickness of the wind launching region is quite small ($\delta R << 0.1$) which does not seem likely (unless the launching region is highly localised such the top point of a warp in the accretion disc), it is difficult to explain all of the results in the context of a wind being launched from close to the inner edge of the disc. We will attempt to pinpoint the launching mechanism of the wind in section \ref{windlaunchingmech}.

Fig. \ref{Distance_Lion} also shows how the inferred distance depends on the ionising luminosity (left sub-plot) and the super-orbital phase (right sub-plot). The distance is strongly anti-correlated with both of these parameters, with one outlier: observation 0673510901. Observation 0783770701 is consistent with a broad range of distances due to the large uncertainty on the wind column density. The fact that observation 0673510901 is an outlier in both plots suggests that the distance does not only depend on the luminosity of the source but also on the super-orbital phase, i.e. we are likely probing different lines of sight from the neutron star (different inclinations and vertical heights from the accretion disc).

We also explore the dependence of the inferred absorption distance on the orbital phase but do not find any noteworthy correlation.

Eq. \eqref{eqdist} can be turned around to extract the minimum density of the outflowing material:

\begin{equation}
\label{eqdens}
n=\frac{N_{\textrm{H}}^{2}\xi}{L_{ion} \delta R^2}
\end{equation}

Minimum densities required are between 10$^{11}$ and 10$^{15}$ cm$^{-3}$ (with most of the measurements falling in the range of  10$^{12}-$10$^{13}$ cm$^{-3}$), and we observe a positive correlation between the density and both the ionising luminosity and the super-orbital phase (opposite trend as the absorption distance, in Fig. \ref{Distance_Lion}). However, the uncertainties in the density measurement are considerably larger than those of the absorption distance due to both $N_{\textrm{H}}$ and $\delta R$ appearing in the second power in Eq. \ref{eqdens}. Finally, we do not observe any significant correlation between the plasma density and the orbital phase of the binary.

\subsection{Mass outflow rate}

\begin{table}
\centering
\caption{Estimates of the maximum distance of absorption and the mass outflow rate of the wind for each observation. The first mass outflow rate estimate (third column) was made using the ionising luminosity, systematic velocity and the ionisation parameter of the material, and the second estimate (fourth column) was calculated using the column density of the material and its systematic velocity.}
\label{Results_Dist_Mass}
\setstretch{1.4}
\begin{tabular}{cccc} 
Obs. ID&Maximum&\multicolumn{2}{|c|}{Mass outflow rate}\\
&distance&$L_{ion}$, $\xi$, $v$&N$_{\textrm{H}}$, $v$\\
&cm&M$_{\odot}~yr^{-1}$&M$_{\odot}~yr^{-1}$ \\
\hline
0134120101 &$ 4.9_{ -3.1 }^{+ 1.7 } \times 10^{8}$&$ 5.0_{ -3.5 }^{+ 3.7 } \times 10^{-9}$&$ 3.5_{ -2.4 }^{+ 1.1 } \times 10^{-6}$ \\
0153950301 &$ 1.2_{ -0.7 }^{+ 0.8 } \times 10^{10}$&$ 1.9_{ -1.6 }^{+ 1.5} \times 10^{-8}$&$ 7.3_{ -6.6 }^{+ 4.9} \times 10^{-7}$ \\
0673510501 &$ 1.6_{ -0.8 }^{+ 0.8 } \times 10^{10}$&$ 9.3_{ -2.2 }^{+ 2.0 } \times 10^{-8}$&$ 1.5_{ -0.8 }^{+ 0.7 } \times 10^{-7}$ \\
0673510601 &$ 1.7_{ -0.4 }^{+ 0.3 } \times 10^{9}$&$ 8.6_{ -3.2 }^{+ 2.9 } \times 10^{-9}$&$ 9.9_{ -3.1 }^{+ 3.2 } \times 10^{-7}$ \\
0673510901 &$ 1.3_{ -0.4 }^{+ 0.5 } \times 10^{11}$&$ 9.8_{ -2.6 }^{+ 2.4 } \times 10^{-8}$&$ 5.4_{ -1.9 }^{+ 1.5 } \times 10^{-8}$ \\
0783770501 &$ 9.5_{ -0.3 }^{+ 0.3 } \times 10^{10}$&$ 1.7_{ -0.4 }^{+ 0.6 } \times 10^{-7}$&$ 1.5_{ -0.7 }^{+ 0.5 } \times 10^{-7}$ \\
0783770601 &$ 5.6_{ -2.2 }^{+ 2.3 } \times 10^{10}$&$ 1.3_{ -0.5 }^{+ 0.3 } \times 10^{-7}$&$ 2.9_{ -1.2 }^{+ 1.2 } \times 10^{-7}$ \\
0783770701 &$ 1.4_{ -1.4 }^{+ 1.0 } \times 10^{11}$&$ 6.0_{ -3.0 }^{+ 3.5 } \times 10^{-8}$&$ 5.1_{ -4.2 }^{+ 16.5 } \times 10^{-8}$ \\
\hline
\end{tabular}
\end{table}

The ionisation level, column density and velocity of the wind can also be used to estimate the mass outflow rate in the wind and compare it with the mass accretion rate. We will use two independent approaches, one making use of the ionisation parameter, the velocity of the outflowing material and the ionising luminosity of Her X-1, and the second one making use of the systematic velocity and the column density of the wind. In both cases we assume that the wind structure is axisymmetric. Again we assume Solar abundances in this section.

First we use the ionisation parameter of gas. The mass outflow rate can be written as:

\begin{equation}
\dot{M}_{out} = \rho A v = ( C_{V} \mu m_{p} n ) ( \frac{\Omega}{4 \pi} 4 \pi R^{2} ) v
\label{eqMout}
\end{equation}

where $\rho$ is the density of outflowing gas, $A$ is the surface area into which it is launched, $v$ the outflow velocity, $C_{V}$ is the volume filling factor (assumed to be constant across different observations), $\mu$ defines the mean atomic mass ($\approx1.2$ for solar abundances), $m_{p}$ the hydrogen ($\approx$ proton) mass, $n$ the ion concentration in the wind, $\frac{\Omega}{4 \pi}$ is the solid angle into which the wind is launched as a fraction of $4 \pi$ and $R$ is the distance from the ionising source at which absorption occurs.

Here we use $v$ as the observed projected velocity, measured from the line blueshift. In principle, if the wind is launched from the disc and is observed close to its launching point (which is probably the case here because Her X-1 is a high inclination system), it will also likely carry a toroidal velocity component from the Keplerian rotation within the accretion disc. Such velocity component is impossible to measure from the absorption line shifts alone. From the absorption distance calculations in Section \ref{Dist_section} (also shown in Fig. \ref{Distance_Lion}), the wind originates at distances of around 10$^5$ R$_{G}$ in the disc, with two exceptions (observations 0134120101 and 0673510601) where it could originate as close as 10$^{3}-$10$^{4}$ R$_{G}$. The corresponding Keplerian velocities at these radii are $\sim$1000 km/s and $\sim$5000 km/s (taking $3 \times 10^{3}$ R$_{G}$), hence the total wind velocity could be significantly larger.

Nevertheless, unless the wind also carries a considerable vertical speed component (which we will argue against later in this work), the toroidal motion will not affect our calculation of the mass outflow rate because the velocity perpendicular to the surface of integration remains the same as the projected component which is observed.

We can now use the definition of the ionising parameter \eqref{eqxi} to show that:

\begin{equation}
\label{eqfinalMout}
\dot{M}_{out} = 4 \pi \mu m_{p} v \frac{L_{ion}}{\xi}  C_{V}  \frac{\Omega}{4 \pi}
\end{equation}

\begin{figure*}
	\includegraphics[width=\textwidth]{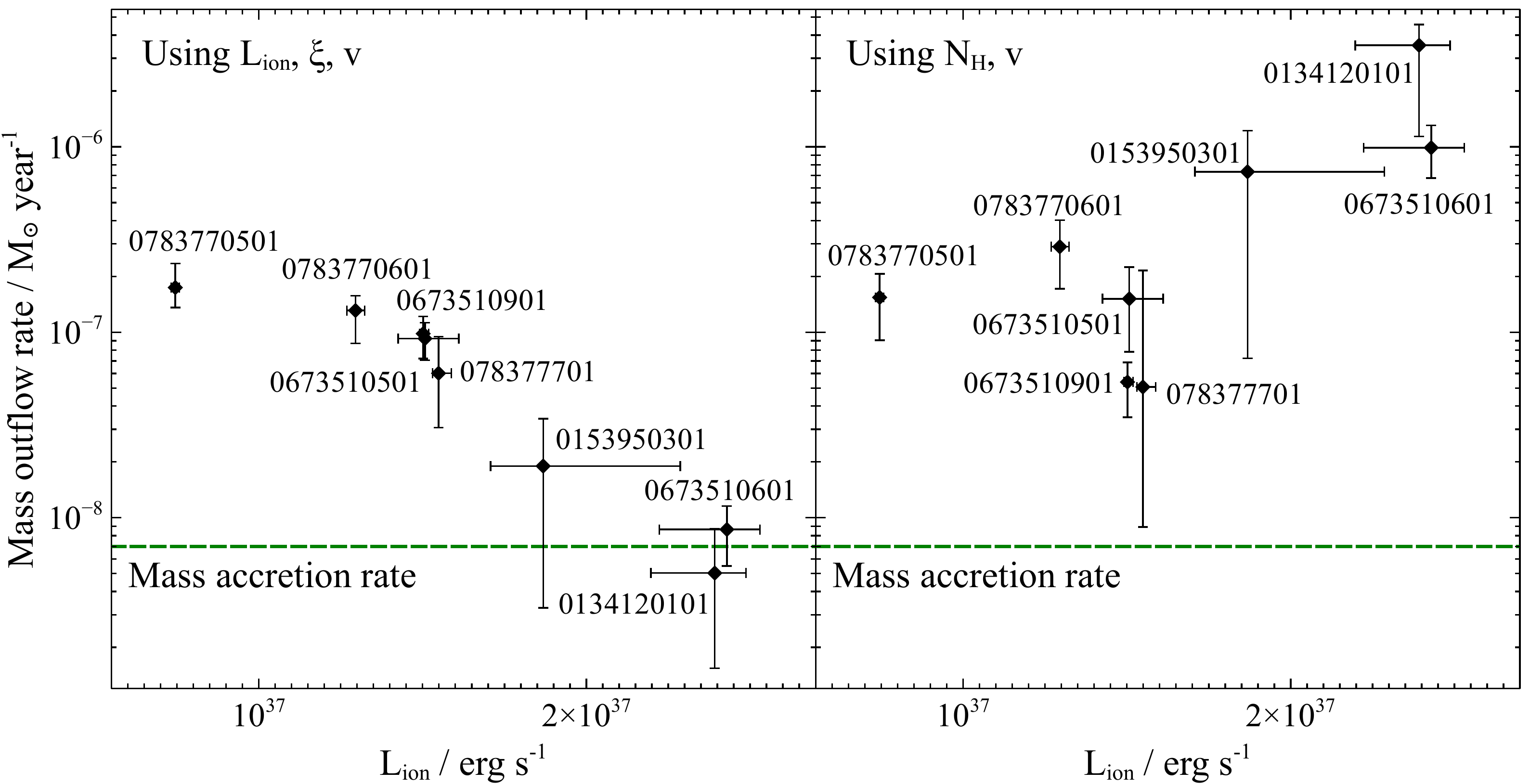}
    \caption{The mass outflow rate of the wind for each observation. The estimate in the left plot was made using the ionising luminosity of the source, the ionisation parameter and the velocity of the material. The estimate in the right plot was made using the column density and the velocity of the wind. The green horizontal lines show the mass accretion rate through the accretion disc \citep{Boroson+07}.}
    \label{Mass_Mass_v_Lion}
\end{figure*}

The volume filling factor $C_{V}$ and the solid angle $\Omega$ are unknown, but we can make an estimate of the mass outflow rate using the remaining parameters (assuming $C_{V}$ and $\Omega$ are equal to unity). The results for each observation are shown in Table \ref{Results_Dist_Mass} and Fig. \ref{Mass_Mass_v_Lion} (left subplot). In reality, both $C_{V}$ and $\Omega$ must be smaller than unity and thus all of the mass outflow rate estimates are upper limits on their real values.

Alternatively, we can estimate the mass outflow rate by assuming that the systematic velocity of the absorber is approximately equal to the escape velocity at the location of wind absorption. Under this assumption, we can write

\begin{equation}
v^{2}= \frac{2GM}{R}
\end{equation}

where $G$ is the gravitational constant and $M$ is the mass of the neutron star. Then, using \eqref{eqnh}, we can express \eqref{eqMout} as

\begin{equation}
\dot{M}_{out} = 4 \pi \mu m_{p} v \frac{N_{\textrm{H}}}{R \delta R}  R^{2} C_{V}  \frac{\Omega}{4 \pi}=4 \pi \mu m_{p} v \frac{N_\textrm{{H}}}{ \delta R}  \frac{2GM}{v^{2}} C_{V}  \frac{\Omega}{4 \pi}
\end{equation}

Therefore we find that

\begin{equation}
\dot{M}_{out} = 4 \pi \mu m_{p} \frac{N_{\textrm{H}}}{\delta R}  \frac{2GM}{v}  C_{V}  \frac{\Omega}{4 \pi} = 8 \pi \mu m_{p} GM \frac{N_{\textrm{H}}}{v} \frac{1}{\delta R}  C_{V}  \frac{\Omega}{4 \pi}
\end{equation}

This assumption necessarily also ignores the projection effect on the actual outflow velocity. The estimated mass outflow rate for each observation is shown again in Table \ref{Results_Dist_Mass} and Fig. \ref{Mass_Mass_v_Lion} (right subplot), taking the mass of the central neutron star to be 1.4 $M_{\odot}$, and assuming the $C_V$, $\Omega$ and $\delta R$ parameters are equal to unity.

We find that most of the mass outflow rate estimates are significantly higher than the mass accretion rate, on average more than an order of magnitude larger. Fig. \ref{Mass_Mass_v_Lion} also shows that the two different mass estimates diverge with luminosity. This suggests that the assumption of the outflow velocity being parabolic is most likely incorrect. This fact is also underlined by the apparent lack of correlations between the outflow velocity and the ionising luminosity, orbital or super-orbital phase. Therefore the mass outflow rates estimated by the second method should be taken cautiously.

There is a strong correlation between the mass outflow rates inferred from the ionisation parameter and the ionising luminosity of the neutron star. This correlation is necessarily driven by the strong dependence of the ionisation parameter on the luminosity.

On the first look, the mass outflow rate estimates appear too high. However, these values do not account for the limited solid angle and volume filling correction factors. I.e., using a modest opening angle of 5$^{\circ}$ (above and below the accretion disc) into which the wind is outflowing, we obtain $\frac{\Omega}{4\pi}\approx0.1$, bringing down the mass outflow rates by an order of magnitude. Nevertheless, even if the volume filling factor is very small as well ($C_{V}\sim0.1$), the mass outflow rates are still of the same order as the mass accretion rates onto the primary and hence the wind is an important component of the accretion flow.

Very high wind mass outflow rates, of the same order as the mass accretion rate or higher, were previously found in other X-ray binaries \citep[e. g.][]{Miller+06, Neilsen+09}. Similar results were also found in studies of warm absorbers \citep{Blustin+05} and ultrafast outflows \citep{Nardini+15} in active galactic nuclei. \citet{Blustin+05} find almost one order of magnitude higher median mass outflow rates than the median mass accretion rates using a sample of 23 AGN with detected warm absorbers.

\subsection{Possible wind launching mechanisms}
\label{windlaunchingmech}

Her X-1 is thought to be consistently in a sub-Eddington mass accretion state with an Eddington ratio of $\sim0.1-0.2$. The hard X-ray radiation from the accretion column is beamed, but not by an extreme factor hence the beamed flux should not reach Eddington levels. Radiation pressure on electrons therefore should not be sufficient to drive the wind from the accretion disc to $\sim$1000 km/s velocities.

Line driving could enhance the pressure on outflowing material, however the material cannot be over-ionised as it is being accelerated away from the disc \citep{Proga+00}. The high inclination of Her X-1 suggests that we might be viewing the wind close to the launching site. The very high ionisation degree ($\log \xi > 3$) therefore implies that the wind is most likely not driven by line pressure.

The wind could originate in the part of the accretion disc periodically irradiated by the beam of the neutron star. This might be the same region that is present in the soft X-ray spectrum of Her X-1, and pulses with the same period but out of phase of the neutron star. The temperature of this component, found in previous studies and also here is approximately 0.1 keV. The sound speed at this temperature is: $c_{s}=\sqrt[]{\frac{kT}{m_{p}}}$, where $k$ is the Boltzmann constant, $T$ the temperature of the plasma and $m_p$ is the hydrogen (proton) mass. For $T=0.1$ keV $ \approx 1.2 \times 10^{6}$ K, we find $c_{s} \approx 100$ km/s, much lower than the outflow velocities observed in this study.

Alternatively, if the hard X-ray radiation illuminates the outer part of the accretion disc (not unlikely as the accretion column radiation is beamed and the disc is warped), these regions could be Compton heated to a high enough velocity to escape the local gravity \citep{Begelman+83}. The `inverse Compton temperature' $T_{IC}$ to which the outer disc regions can be heated to is defined as 

\begin{equation}
k T_{IC} = \frac{1}{4} \langle \epsilon \rangle = \frac{1}{4} L^{-1} \int^{\infty}_{0} E L_{E} dE
\end{equation}

where $L$ is the total luminosity and $L_{E}$ is the luminosity at energy $E$. $T_{IC}$ for each of our observations of Her X-1 vary, and also depend on whether the full X-ray SED is taken or only the accretion column radiation (the \textsc{comt} component, producing harder radiation than the full SED). However, all the estimates are in the range between 2.5 and 3.5 keV, and on the upper end of this interval if only the accretion column radiation is taken as the heat source. The $T_{IC}$ values for each observation are listed in Table \ref{Results_IC}.

\begin{table}

\centering
\caption{Inverse Compton temperatures and approximate wind launching radii for each observation. The second and third columns contain values calculated using the full SED of Her X-1, the fourth and fifth columns show the values calculated using just the accretion column radiation as the SED.}
\label{Results_IC}
\setstretch{1.4}
\begin{tabular}{ccccc} 
Obs. ID&\multicolumn{2}{|c|}{Full SED}&\multicolumn{2}{|c|}{Accretion column SED}\\
&T$_{\textrm{IC}}$&0.1$\times$R$_{\textrm{IC}}$&T$_{\textrm{IC}}$&0.1$\times$R$_{\textrm{IC}}$\\
&keV&cm&keV&cm\\
\hline

0134120101&2.21 &1.05$\times10^{10}$ &2.67 &8.72$\times10^{9}$ \\
0153950301&2.24 &1.04$\times10^{10}$ &2.76 &8.44$\times10^{9}$ \\
0673510501&2.36 &9.87$\times10^{9}$ &2.88 &8.09$\times10^{9}$ \\
0673510601&2.64&8.82$\times10^{9}$ &3.18 &7.33$\times10^{9}$  \\
0673510801&2.79 &8.35$\times10^{9}$ &3.26 &7.15$\times10^{9}$  \\
0673510901&2.58& 9.03$\times10^{9}$&3.01 &7.74$\times10^{9}$  \\
0783770501&2.39 &9.75$\times10^{9}$ &2.65 &8.79$\times10^{9}$ \\
0783770601&2.58 &9.03$\times10^{9}$ &3.04 &7.66$\times10^{9}$  \\
0783770701&2.62 &8.89$\times10^{9}$ &3.05 &7.64$\times10^{9}$  \\

\hline
\end{tabular}
\end{table}

Thus a Compton-heated thermal wind will occur for accretion disc radii greater than \citep[using equations 2.7 and 2.8 from][]{Begelman+83}

\begin{equation}
R/R_{IC} \gtrsim 0.1
\end{equation}

where R$_{IC}$ is:

\begin{equation}
R_{IC} = \frac{GM \mu m_{p}}{kT_{IC}}
\end{equation}

If we take $kT_{IC} \sim 3$ keV $\sim 3.5 \times 10^{7}$ K, the wind should be launched at distances larger than $R \sim 8 \times 10^{9}$ cm (4$\times$10$^{4}$ R$_{\textrm{G}}$) from the neutron star (precise values for each observation are tabulated in Table \ref{Results_IC}). This corresponds to the outer half of the accretion disc, and matches very well with most of our estimates of the absorption distance from the ionising source (Section \ref{Dist_section}). At the same time, the critical luminosity required for a significant mass flux within the wind is \citep[eq. 2.12b from][]{Begelman+83}

\begin{equation}
L_{cr} = 0.030~(T_{IC}/10^8 K)^{-1/2}~L_{\epsilon}
\end{equation}

where $L_{\epsilon}$ is the Eddington luminosity:

\begin{equation}
L_{\epsilon} = 1.5 \times 10^{38} (M/M_{\odot})$ erg/s$
\end{equation}

For $T_{IC} \sim 3.5 \times 10^{7}$ K, we obtain $L_{cr} \sim 1.1 \times 10^{37}$ erg/s. This is just about the luminosity of the lowest flux observation (0783770501, Table \ref{Results}) and much lower than the average flux during remaining observations. Thus the luminosity of Her X-1 is above the critical luminosity and a Compton-heated thermal wind should occur in the high state of Her X-1 with a significant mass outflow rate if the outer part of the accretion disc is illuminated by the accretion beam. Unfortunately, we cannot estimate the mass outflow rate from theory as the value depends on the pressure of gas in the launching point of the wind. 

On the other hand, we can roughly estimate the outflow velocity. The rms velocity of a Compton-heated particle of the wind ($T_{IC} \sim 3$ keV) is:

\begin{equation}
v = \sqrt{\frac{3kT}{\mu m_{p}}} \sim 850 $km/s$
\end{equation}

The outflow velocity estimate compares very well to the values we are measuring in most observations. Compton heating is therefore a very plausible candidate for driving the wind.

Lastly, magnetic driving is also a possible launching mechanism for the wind \citep{Blandford+82}. Magnetic field of the neutron star \citep[$B_0\approx4\times 10^{12}$ G on the star surface,][]{Staubert+19} could potentially provide the driving force. If the neutron star field is a dipole, at distance $R$ it will decay such as $B=B_{0}(R/R_{0})^{-3}$ where $R_0$ is the neutron star radius (the strength will decay faster if the field is higher order than a dipole). Thus the field strength at the inner accretion disc radius ($\sim10^9$ cm) is roughly $10^3-10^4$ G, which could be sufficient to drive the wind \citep{Ustyugova+99,Donati+07,Miller+16b}. If the outflow however originates beyond $R\sim10^{10}$ cm, which is likely the case for most of our \xmm\ observations, the field will be of order of $\sim10$ Gauss or less, insufficient to launch the material. Alternatively, additional magnetic field could originate in the accretion disc itself. However, as we cannot readily estimate this field, it is difficult to judge whether it can or cannot contribute to or drive the wind in this object.

Winds with similar properties (velocities, ionisation parameters) have been discovered in other neutron star binary systems such as GX 13+1 \citep{Ueda+04}, IGR J17480-2446 \citep{Miller+11} and IGR J17591-2342 \citep{Nowak+19}. The difference between these three and Her X-1 is that the column density observed in Her X-1 is significantly lower during most observations and thus the absorption strength is correspondingly lower. The outflow properties are also comparable to the ionised winds observed in Galactic black hole binaries such as GRO J1655-40 \citep{Miller+06, Fukumura+17}, GRS 1915+105 \citep{Neilsen+09} and other objects where both Compton thermal and magnetic driving were invoked to explain the wind origin.

On the other hand, one important difference between all of the aforementioned examples is that Her X-1 is known to be a highly magnetised ($10^{12}$ G) neutron star. To our current knowledge, Her X-1 is the first highly magnetised neutron star to harbour such a wind. The magnetic field effectively puts an absolute lower limit of $\sim1000~$R$_{\textrm{G}}$ on the wind launching radius as it truncates the inner accretion disc. This can be used as an effective constraint in theoretical models explaining the appearance of the wind.

\subsection{Are observations over the super-orbital period probing different lines of sight?}

We observe an inverse correlation between the distance at which the wind absorption occurs and the ionising luminosity of Her X-1 (Fig. \ref{Distance_Lion}). The relation could suggest that the wind launching radius decreases with increasing luminosity. Such a correlation would be expected for example if the radiation pressure on the outflowing material played a role. This explanation would come naturally within the framework of a wind driven by a combination of Compton heating and radiative force.

At the same time, the correlation between the distance and luminosity necessarily implies that the absorption distance is correlated with the super-orbital phase too (since luminosity and super-orbital phases are connected, from Fig. \ref{Lion_suporb}). If the super-orbital phase is the driver for the correlation instead of the ionising luminosity, an alternative explanation for the variations is possible. The super-orbital period is likely introduced by the precession of a warped accretion disc, which at times obscures our view into the innermost accretion disc and onto the neutron star. But even in the high state, when our view of the inner regions is unobstructed, the disc continues precessing and the angle between different annuli of the disc and our line of sight keeps changing \citep[Fig. 3 of ][]{Leahy+02}. It is thus possible, that by observing the wind at different super-orbital phases, we are probing different heights of the wind above the accretion disc. 

\begin{figure}
	\includegraphics[width=\columnwidth]{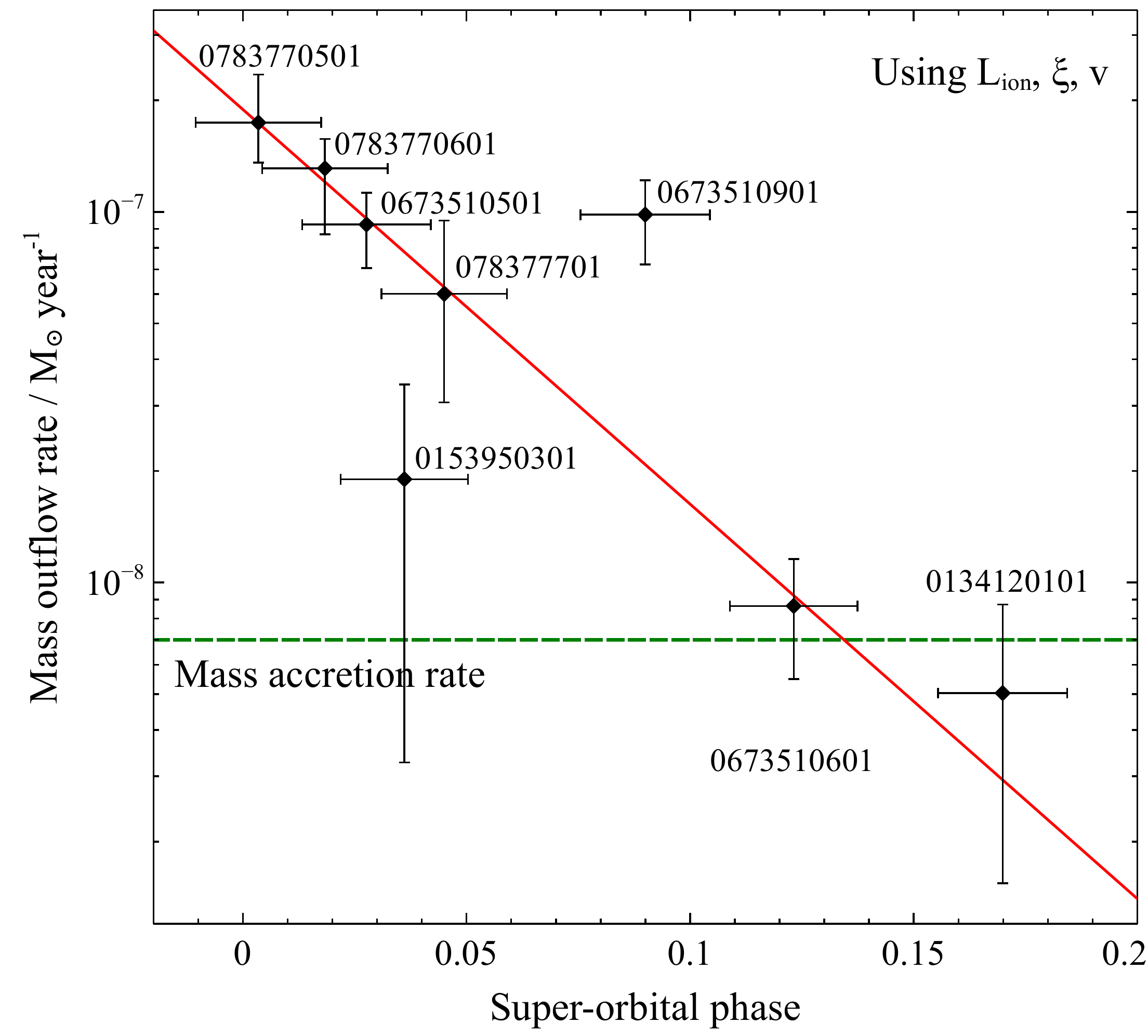}
    \caption{Wind mass outflow rate versus the super-orbital phase of each observation. The red line shows the best log-linear fit to the data, excluding observations 0153950301 and 0673510901 and treating them as outliers.}
    \label{Mass_suporb}
\end{figure}

If this is the case, we are presented with a rare opportunity to study the vertical structure of a disc wind. We might be able to determine the opening angle of the outflow (except for any wind component which is fully ionised) and consequently obtain an accurate estimate of the wind mass outflow rate. Fig. \ref{Mass_suporb} shows that at the highest ionising luminosities (i.e. at super-orbital phases of 0.1-0.2), the mass outflow rate estimates are an order of magnitude smaller than those at super-orbital phase equal to 0 (the turn-on point).

At this point we can make a zeroth-order effort to calculate the real mass outflow rate which accounts for the limited wind launching solid angle. We can use the fact that the mass outflow rate decreases very steeply with the increasing super-orbital phase. We  assume that each super-orbital phase corresponds (`maps to') to a specific angle between our line of sight and the accretion disc plane. We also assume that the dependence between this angle and the phase is linear. This is naturally a gross oversimplification, and in reality the function might resemble a sinusoid \citep[but not necessarily, we again refer to Fig. 3 from][]{Leahy+02}. At phase equal to 0, the angle is also zero - the surface of the accretion disc is just about grazing our line of sight towards the neutron star. At the highest luminosity of Her X-1, which occurs roughly around phase equal to 0.12 \citep{Staubert+13}, the angle will then also reach a maximum.

We will further assume that this inclination is the maximum angle between our line of sight towards the neutron star and the accretion disc. The maximum angle is likely around $\theta \approx5^\circ\approx0.087$ rad \citep{Leahy+02}. Then the mass outflow rate per solid angle can be obtained from eq. \ref{eqfinalMout} as: 

\begin{equation}
\frac{d\dot{M}_{out}}{d\Omega} = \frac{\dot{M}_{exp}}{4\pi}
\end{equation}

where $d\dot{M}_{exp}$ is the experimentally measured mass outflow rate value which assumes a full solid angle and full volume filling factor (this is the value listed in Fig. \ref{Mass_Mass_v_Lion} and Table \ref{Results_Dist_Mass}). We can relate the solid angle to an inclination angle $\theta$ from the disc such as:

\begin{equation}
d\Omega=\frac{dA}{R^2}=\frac{2 \times 2 \pi R^{2}\cos {\theta} d\theta}{R^2} \approx 4 \pi d\theta
\end{equation}

for very small angles $\theta$ (satisfied here). The first factor 2 in the numerator of the fraction is introduced because the disc wind is launched from both sides of the accretion disc. Then the total mass outflow rate can be integrated as: 

\begin{equation}
\dot{M}_{out} = \int_{}^{} \frac{\dot{M}_{exp}}{4 \pi} d\Omega=\int_{}^{} \frac{\dot{M}_{exp}}{4\pi} 4\pi d\theta = \frac{d\theta}{d\phi} \int_{}^{} \frac{\dot{M}_{exp}}{d\Omega} d\phi 
\end{equation}

where $\phi$ is the super-orbital phase. Now taking $\phi_{max}\approx0.12$, $\theta_{max}\approx0.087$ rad, if $\phi$ and $\theta$ are linearly related (as assumed) we obtain $\frac{d\theta}{d\phi}=0.73$. It remains to integrate $\int_{}^{} \frac{d\dot{M}_{out}}{d\Omega} d\phi$. Fig. \ref{Mass_suporb} shows the best log-linear fit to measured mass outflow rates versus the super-orbital phase, excluding observations 0153950301 and 0673510901 which appear to be outliers. The best-fitting relation is in form

\begin{equation}
\log (\dot{M}_{out}/M_{\odot}~$year$^{-1})=-15.48-24.51\times\phi
\end{equation}

We integrate from phase $\phi_{min}=0$ to $\phi_{max}=0.12$, obtaining $7.3\times10^{-9}~M_{\odot}~$year$^{-1}$. If we use all of the observations in the calculation (including the apparent outliers), we instead get $5.8\times10^{-9}~M_{\odot}~$year$^{-1}$. With this approach we are ignoring any wind at inclination angles higher than $\theta_{max}\sim5^\circ$, which never crosses our line of sight. However, the mass outflow rate seems to be a strongly decreasing function with super-orbital phase (and consequently the inclination angle) and therefore the error on the total mass outflow rate introduced this way should be small. The total mass outflow rate is: 

\begin{equation}
\dot{M}_{out} \approx \frac{d\theta}{d\phi} \int_{}^{} \frac{d\dot{M}_{out}}{d\Omega} d\phi \approx 5\times10^{-9}~M_{\odot}~$year$^{-1}
\end{equation}

Alternatively, if all of the observations are included in the calculation, we obtain $\sim4\times10^{-9}~M_{\odot}~$year$^{-1}$. Both of these estimates are very comparable to the mass accretion rate which is around $7\times10^{-9}~M_{\odot}~$year$^{-1}$ \citep[measured from the UV spectrum of the accretion disc,][]{Boroson+07}. Finally, we note that this estimate still does not account for the volume filling factor of the wind.

If the super-orbital disc precession drives the wind variation, the outflow must have a small solid launching angle with an opening angle of $\sim5^{\circ}$. It is possible that fully ionised material, undetectable with X-ray spectroscopy, is launched to larger solid angles. However, at high ionisation levels this material contributes little to the overall mass outflow rate budget, unless it is launched at much larger velocities. A small wind opening angle is also supported by the fact that any P-Cygni profiles from outflowing material not along our line of sight are weak or non-existent \citep[as opposed to other binaries where evidence of P-Cygni features was seen, ][]{Miller+15b}. This is especially the case for the highest quality observations (0673510501 and 0783770601). Furthermore, at higher super-orbital phases, we do not observe any signatures of winds from phases close to 0, which suggests that the wind is launched directly at us, i.e. its vertical velocity component is likely not large.

At the moment it is not clear which of the above possibilities is correct - whether the evolution of the wind is dictated by the ionising luminosity, by the super-orbital phase or in fact by some combination of both. Future studies, which will sample wind evolution during the super-orbital phase more densely, might be able to answer this question. To conclude, Hercules X-1 is a fascinating system that potentially offers us a unique opportunity to sample the vertical structure of an equatorial accretion disc wind.

\section{Conclusions}

We have performed an in-depth X-ray spectral analysis of Hercules X-1 with \xmm\ combining both broadband and high-spectral resolution instruments and focusing mainly on the high flux state of this famous object. The conclusions of our study are follow:

\begin{itemize}

\item We detect a highly ionised blueshifted wind in the high state of Her X-1. The detection is statistically significant in most of the \xmm\ observations. The wind has a projected outflow velocity of 200 to 1000 km/s, varying between different observations, and an ionisation parameter in the range of $\log \xi$ of 3.0 to 5.0. Her X-1 can thus be added to a list of known neutron star binaries with detected ionised outflows, such as GX 13+1, IGR J17480-2446 and IGR J17591-2342.

\item While photon pile-up due to very high X-ray fluxes might affect a fraction of our datasets, especially pn and RGS 2 data, we conclude that the wind detection is robust and the absorption features are not induced by this instrumental effect.

\item We find a clear correlation between the luminosity of Her X-1 and the ionisation parameter of the wind, suggesting that the material in the outflow observes a similar X-ray flux as we do. We also observe a correlation between the super-orbital phase and the ionisation parameter. We do not observe any significant correlations between the outflow velocity and these two parameters.

\item The best-fitting wind properties suggest that the absorption takes place within the accretion disc of the primary. It is plausible that the X-ray outflow, launched from the accretion disc, is the progenitor of the UV blueshifted absorption observed in Her X-1, but the latter only occurs at larger, circumbinary distances from the system.

\item We find that the mass outflow rate in the wind can easily be of the same order as the mass accretion onto Her X-1. It is thus an important component of the accretion process. We conclude that it is driven either by Compton heating of the outer accretion disc or by magnetic fields.

\item If the super-orbital phase is the driver for the wind variability instead of the ionising luminosity, the individual observations are likely observing different lines of sight above the accretion disc and sampling the vertical structure of the wind. If this is the case, we can estimate the mass outflow rate corrected by the limited launching solid angle to be $5\times10^{-9}~M_{\odot}~\textrm{year}^{-1}$, approximately 70\% of the mass accretion rate.

\item Our chemical analysis of elemental abundances in the ionised wind finds a strong over-abundance of iron compared to oxygen. We also confirm results of previous studies which found an over-abundance of nitrogen and neon compared to oxygen. At the same time, there is evidence that oxygen is not in fact under-abundant compared to the remaining chemical elements.

\item None of the low state or short-on \xmm\ observations of Her X-1 are of high enough quality to detect a similarly ionised wind. We stack all of the available low state data but do not find obvious evidence of blueshifted absorption. Future long exposure observations, or observations with instruments with higher collecting area \citep[such as Athena,][]{Nandra+13} might be able to detect a wind if present.

\end{itemize}

\section*{Acknowledgements}

PK acknowledges support from the STFC. ACF acknowledges support from ERC Advanced Grant Feedback 340442. CP is supported by European Space Agency (ESA) Research Fellowships. DJW acknowledges support from STFC Ernest Rutherford fellowships. CSR thanks the UK Science and Technology Facilities Council for support under Consolidated Grant ST/R000867/1. This work is based on observations obtained with XMM-Newton, an ESA science mission funded by ESA Member States and USA (NASA).





\bibliographystyle{mnras}
\bibliography{References} 




\appendix

%

\section{Pile-up}

\label{Appendix}

\begin{figure*}
	\includegraphics[width=\textwidth]{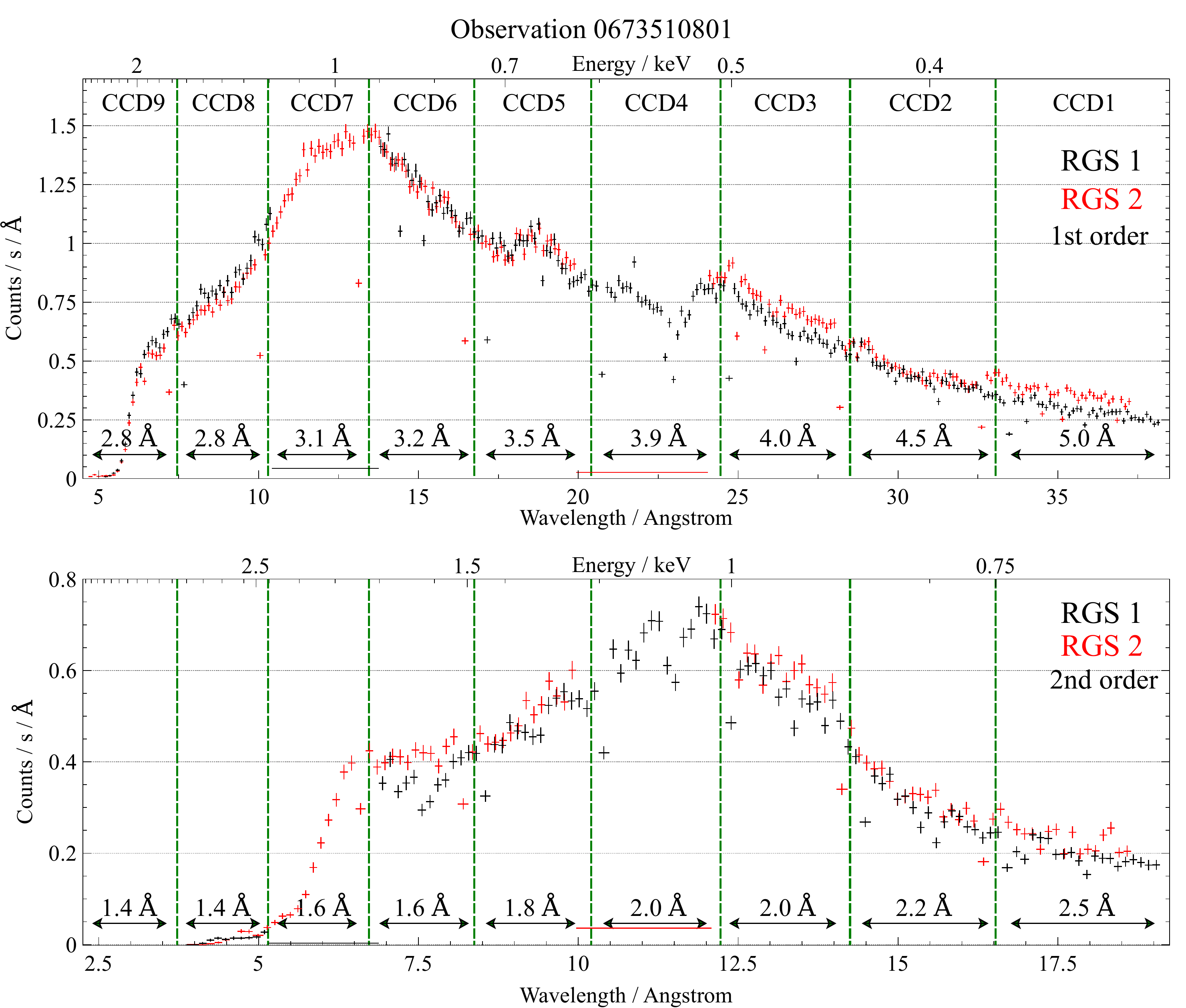}
    \caption{The average count rate per \AA\ in the first and the second order of both RGS 1 and 2 instruments during the highest count rate observation 0673510801. The top plot contains the first order count rates and the bottom plot contains the second order count rates. RGS 1 data are in black, while RGS 2 data are red. Green vertical dashed lines show the approximate positions of chip gaps and the labels list the approximate chip sizes in \AA. CCD numbers are listed on the top of the figure. The count rate per CCD can be approximated as the sum of the count rates per \AA\ times the CCD width (in \AA) of both orders.}
    \label{Pileup_0673510801}
\end{figure*}

One important issue which needs to be discussed is that of pile-up. Table \ref{Obsdata} shows that the count rates in both RGS and pn during most of the observations are indeed very high. Count rates of up to $\sim$800 counts/s in pn are reached during some of the observations, which is close to the pile-up limit for timing mode observations. Furthermore, this does not take into account any variability and especially pulsations of the source. It is therefore not surprising to expect some pile-up in the higher flux observations.

The count rates observed in the RGS cameras are also high, with up to about 20 counts/s per camera. The pile-up threshold in RGS is usually defined by a maximum count rate per CCD in RGS 1 or 2. Its effect is to move first order events to the second order and thus degrade both of these spectra. The threshold is roughly 12 counts/s per CCD in RGS 1 and 6 counts/s per CCD in RGS2 (because one of the two read-out nodes in RGS 2 is disabled) \footnote{www.cosmos.esa.int/web/xmm-newton/sas-thread-pile-up-in-the-rgs}. This is the maximum summed count rate of all orders per CCD. 

Fig. \ref{Pileup_0673510801} shows the count rate per \AA ngstrom for both orders and RGS detectors in observation 0673510801, the observation with the highest RGS flux. The figure also shows the approximate CCD edge positions and the width of each chip in \AA. The total count rate per CCD can be calculated as the sum of the count rates per \AA\ for each order times the width of each CCD in \AA\ (which is different in each order, the width in the second order is half that of the first order). The figure shows that multiple CCDs between CCD3 and CCD7 could be affected by pile-up. The approximate fluxes per CCD are: CCD 8 - 2.5 cts/s, CCD 7 - 4.4 cts/s, CCD6 - 4.6 cts/s, CCD 5 - 4.5 cts/s, CCD 4 - 4.5 cts/s, CCD 3 -4.2 cts/s and CCD 2 - 2.8 cts/s. These values are hence below the 6 cts/s RGS 2 limit, but again, the are only the average count rates for a variable and pulsating source. It is not inconceivable to imagine a 50 per cent flux spike (e.g. during the pulse) during which the count rate in some CCDs would exceed 6 cts/s. We thus conclude that while RGS 1 should be unaffected by pile-up, RGS 2 might be partially affected in the highest flux observations (but probably less than pn).

It is unlikely that pile-up could introduce a series of narrow absorption features in both the pn and the RGS spectra that line-up in velocity space. Nevertheless, we perform two checks to show that the wind features are indeed real in at least some of the observations in this study. First, we exclude pn data from the analysis and only look at RGS1 and RGS 2 (first order) data, performing a similar fit as described in \ref{high_state}. Secondly, we also exclude RGS 2 from the analysis and only use RGS 1 first and second order data to show that the wind is still significantly present at least in one observation with the highest data quality.

\subsection{Analysis with RGS 1 and 2}

\begin{table*}
\centering
\caption{Best-fitting wind parameters for each observation of Her X-1 using only RGS 1 and RGS 2 instruments. The first column contains the observational ID. The remaining columns show the properties of outflowing wind such as its column density, ionisation parameter, turbulent velocity and systematic velocity, as well as the statistical fit improvement of the final model compared to the baseline continuum spectral model.}
\label{Results_RGS}
\setstretch{1.4}
\begin{tabular}{cccccccc} 
Obs. ID&Column&log $\xi$&Turbulent &Outflow &$\Delta$C-stat\\
&density&&velocity&velocity& \\
	&	10$^{24}~$cm$^{-2}$	&	erg~cm~s$^{-1}$		&	km$~$s$^{-1}$	&		km$~$s$^{-1}$& \\
\hline
0673510501&$3.04 \pm 0.1$2& $5.05_{-0.10}^{+0.06}$&$250 \pm 170$&$-940 \pm 130$&40.89\\
0673510601&$0.17_{-0.08}^{+0.34}$&$4.79_{-0.12}^{+0.14}$&$30_{-30}^{+200}$&$-330 \pm 150$ &8.02\\
0673510901&$0.11_{-0.07}^{+0.14}$&$4.28_{-0.06}^{+0.08}$&$160_{-100}^{+130}$&$-610_{-130}^{+140}$&23.46\\
0783770501&$0.0026_{-0.0013}^{+0.0023}$&$3.01_{-0.13}^{+0.12}$&$150_{-80}^{+140}$&$-730_{-170}^{+210}$&14.17\\
0783770601&$2.4_{-0.8}^{+1.5}$&$4.54\pm0.12$&$80_{-50}^{+30}$&$-380_{-100}^{+90}$&54.79\\
\hline
\end{tabular}
	\\
\end{table*}

The spectral model has to be modified after ignoring the pn data. The hard band (3-10 keV) data are unavailable and hence the overall broadband spectral shape cannot be constrained. We thus replace the \textsc{comt} Comptonisation continuum model with a simpler \textsc{powerlaw}. Naturally the 6.6 keV Gaussian line is omitted as it is out of the current energy band (0.35-1.8 keV). The remaining model components are identical as before. The continuum model is therefore: \textsc{hot$\times$(powerlaw+bb+4$\times$gauss+cie)}. Since we do not have access to a broadband 0.3-10 keV SED while only using the RGS data, we also have to resort to using the \textsc{xabs} photoionisation absorption model with the default SED.

The data is first fitted with the continuum model, after which we add the wind absorption. We fit for the wind parameters and also recover the fit improvement. The results are listed in Table \ref{Results_RGS}. For this check we only use the 5 observations with the strongest significance of the wind detection. It is evident that while for some observations, there is a large decrease of significance after ignoring the pn dataset, multiple still show strong evidence for the wind absorption. Most importantly, observations 0673510501 and 0783770601 show a $\Delta$C-stat of more than 40 which confirms the reality of the absorption features. 

We also notice (Table \ref{Results_RGS}) that the RGS only analysis predicts very high ionisation levels of the outflowing gas, much higher than the full broadband analysis. This is because of the absence of the crucial Fe K band containing the \ion{Fe}{XXV} and \ion{Fe}{XXVI} ions. These transitions are necessary for putting upper limits to the ionisation parameter of plasma at high ionisation levels.

\subsection{Analysis with RGS 1 only}

The most stringent test is to only use RGS 1 data, being the instrument that is the least affected by pile-up. Unfortunately, CCD 7 is disabled in RGS 1 so  important $10-13$ \AA\ waveband containing the \ion{Ne}{X} absorption line is lost. We therefore also use the second order RGS 1 data, which has access to this part of the spectrum, although we note that the data quality is significantly lower.

\begin{figure}
	\includegraphics[width=\columnwidth]{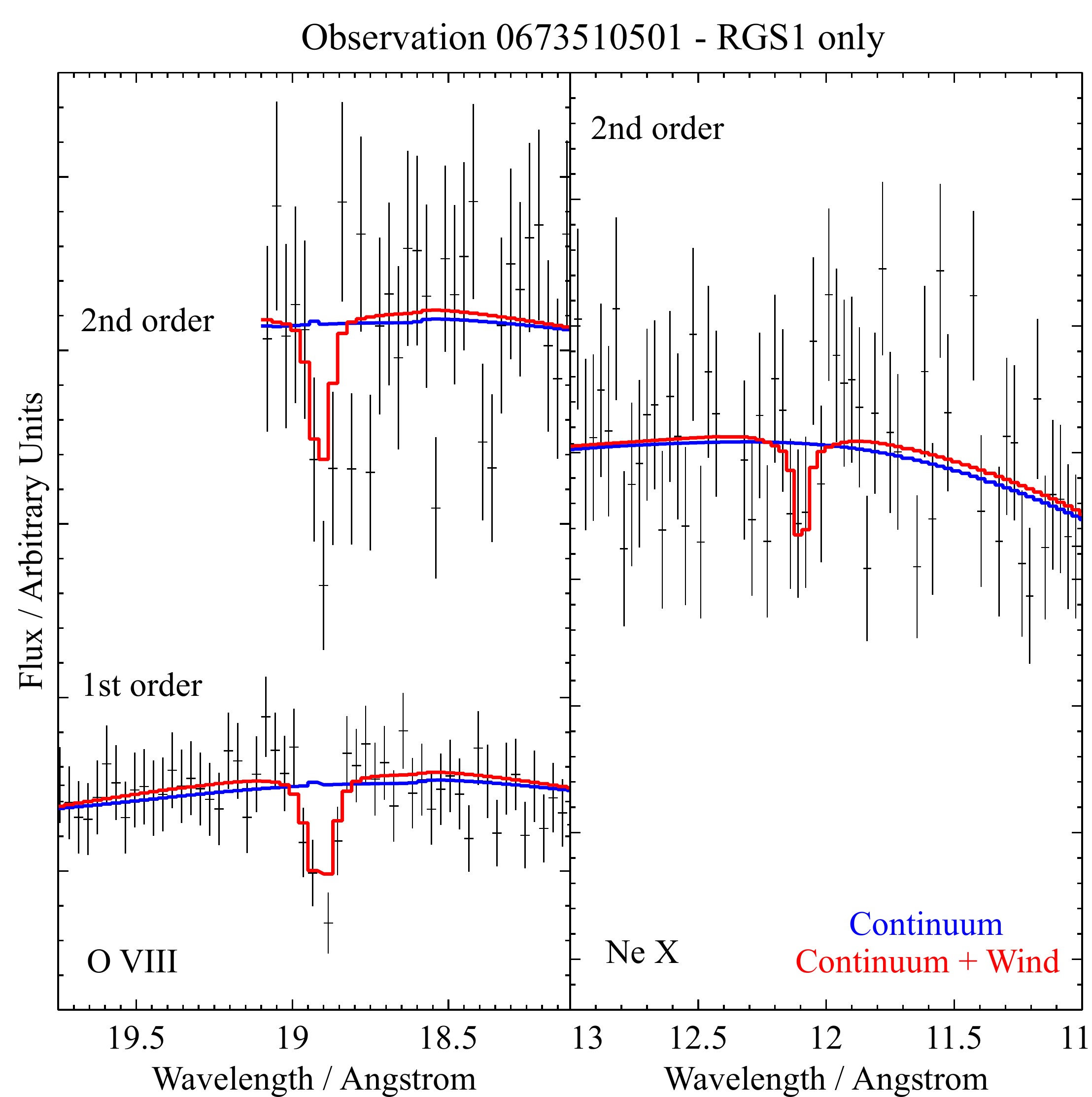}
    \caption{The best-fitting spectral models for observation 0673510501 using only RGS 1 first and second order data. Only the \ion{O}{VIII} and \ion{Ne}{X} bands are shown as the \ion{N}{VII} signature is insignificant in this case. The best-fitting baseline continuum is shown in blue and the final wind solution in red.}
    \label{Spectrum_0673510501_RGS1}
\end{figure}

Here we only use data from observation 0673510501 (which has the strongest wind features) to show that even in the worst case of pile-up, there is at least one observation with a statistically significant wind detection. We use the same continuum and wind spectral models as in the previous section. The best-fitting spectra are shown in Fig. \ref{Spectrum_0673510501_RGS1}. The recovered wind parameters are: column density of $(0.16\pm0.04)\times10^{24}~$cm$^{-2}$, ionisation parameter of $\log \xi=4.2\pm0.2$, systematic velocity of $-1140\pm120~$km/s and a turbulent velocity of $280_{-90}^{+120}~$km/s. The fit improvement $\Delta$C-stat is 33.12. This is a large enough fit improvement to consider the wind detection significant (although we note again that extensive Monte Carlo simulations would be necessary to prove it rigorously).

\bsp	
\label{lastpage}
\end{document}